%% file: astroph.tex
\documentclass{aa} 
\usepackage{graphicx}
\usepackage{txfonts}
\usepackage{xcolor}
\usepackage{natbib}

\newcommand{\Lx}{\mbox{${L_{\mathrm X}}$}}

\newcommand{\fbol}{\mbox{${f_{\rm bol}}$}}

\newcommand{\Rv}{\mbox{${R_{\rm V}}$}}
\newcommand{\Av}{\mbox{${A_{\rm V}}$}}
\newcommand{\EBV}{\mbox{$E_{B-V}$}}
\newcommand{\Ak}{\mbox{$\mathrm{A_{K}}$}}

\newcommand{\Mv}{\mbox{$\mathrm{M_{V}}$}}
\newcommand{\Mbol}{\mbox{$M_{\rm bol}$}}
\newcommand{\kms}{\mbox{$\mathrm{km\,s}^{-1}$}}
\newcommand{\ergcms}{\mbox{$\mathrm{erg\,cm^{-2}\,s^{-1}}$}}
\newcommand{\ergs}{\mbox{$\mathrm{erg\,s^{-1}}$}}
\newcommand{\Nh}{\mbox{$N_{\mathrm H}$}}
\newcommand{\Ks}{\mbox{$K_{\rm s}$}}

\newcommand{\Msun}{\mbox{${M}_{\odot}$}}

\newcommand{\Teff}{\mbox{${T_{\rm eff}}$}}
\newcommand{\Lbol}{\mbox{${L_{\rm bol}}$}}

\newcommand{\Vinf}{\mbox{$\varv_\infty$}}
\newcommand{\LxLbol}{\mbox{$\log({L_{\rm X}}/{L_{\rm bol}})$}}
\newcommand{\fxfbol}{\mbox{$\log({f_{\rm X}}/{f_{\rm bol}})$}}
\newcommand{\logfbol}{\mbox{$\log({f_{\rm bol}})$}}
\newcommand{\logfx}{\mbox{$\log({f_{\rm X}})$}}
\newcommand{\logLx}{\mbox{$\log({L_{\rm X}})$}}
\newcommand{\logg}{\mbox{${\log(g)}$}}
\newcommand{\Mdot}{\mbox{${\dot M}$}}
\newcommand{\Lkin}{\mbox{${L_{\rm kin}}$}}
\newcommand{\xmm}{{\em XMM-Newton}}
\newcommand{\cxo}{{\em Chandra}}
\newcommand{\xo}{ GOSXMM}

\begin{document}

\title{The X-ray catalog of spectroscopically identified Galactic O stars}
\subtitle{Investigating the dependence of X-ray luminosity on stellar and wind 
parameters }
\titlerunning{X-ray catalog of spectroscopically identified Galactic O stars}
\author{Nebot G\'omez-Mor\'an, A.
       \inst{1}
       \and
       Oskinova L.\,M.\inst{2,3}
       }

\institute{Observatoire Astronomique de Strasbourg, Universit\'e de Strasbourg,
     CNRS, UMR 7550, 11 rue de l'Universit\'e, F-67000 Strasbourg, France\\
   \email{ada.nebot@astro.unistra.fr}
   \and
         Institute of Physics and Astronomy, University of Potsdam, 14476, Potsdam,
         Germany \\
         \email{lida@astro.physik.uni-potsdam.de}
    \and
   Department of Astronomy, Kazan Federal University, Kremlevskaya Str 18, 
Kazan, Russia\\}

\date{} 

  \abstract
{
The X-ray emission of O-type stars was first discovered in the early days of the 
{\em Einstein} satellite. Since then many different surveys have confirmed  
that the ratio of X-ray to bolometric luminosity in O-type stars is roughly 
constant, but there is a paucity of studies that account for   
detailed information on spectral and wind properties of O-stars.   
Recently a significant sample of O stars within our Galaxy was 
spectroscopically identified and presented in the 
Galactic O-Star Spectroscopic Survey (GOSS). At the same time, 
a large high-fidelity catalog of X-ray sources detected by  
the \xmm\ X-ray telescope was released.
Here we present the X-ray catalog of O stars with known spectral types
and investigate the dependence of their X-ray properties on spectral type as well as 
stellar and wind parameters. 
We find that, among the GOSS sample, 127 O-stars have a unique \xmm\ source
counterpart and a Gaia data release 2 (DR2) association. Terminal velocities
are known for a subsample of 35 of these stars. We confirm that the X-ray luminosities
of dwarf and giant O stars correlate with their bolometric luminosity. For the subsample of O stars with
  measure terminal velocities we find that
the X-ray luminosities of dwarf and giant O stars also correlate with wind parameters.
However, we find that these correlations break down for supergiant stars. 
Moreover, we show that supergiant stars are systematically harder in X-rays 
compared to giant and dwarf O-type stars. We 
find that the X-ray luminosity 
depends on spectral type, but seems to be independent of whether the stars are 
single or in a binary system. Finally, we show that the distribution of 
\LxLbol\ in our sample stars is non-Gaussian, with the peak of the 
distribution at $\LxLbol\approx -6.6$.
}

\maketitle
%

\section{Introduction}
Since the era of the {\em Einstein} X-ray telescope (0.2--4.0\,keV) 
it is known that O-type  stars emit X-rays \citep{harndenetal79-1}. Based on 
the initial sample of 16 OB stars detected by {\em Einstein}, 
\cite{long+white80-1} noted that their X-ray luminosity  
correlates with bolometric luminosity
as $\Lx\sim 10^{-6\,...\,-8}\Lbol$. 
\cite{pallavicinietal81-1} extended the study to 35 stars with spectral types 
in the range from O4 to A9  and concluded that the ratio of X-ray to bolometric 
luminosity is roughly constant ($\Lx\approx 
(1.4\pm0.3)\times10^{-7}\Lbol$), breaking down for spectral types later 
than A5. Later on, \cite{schmittetal85-1} showed that the $\LxLbol\sim -7$ 
correlation  does not  hold for  A-type stars breaking down already at 
spectral type B5.
\cite{chlebowski+harnden89-1} presented ``The {\em Einstein} 
X-ray Observatory Catalog of O-type stars''.  The catalog contains 289 stars 
with 89 detections and 176 upper limits. It was found that X-ray luminosities of 
O stars are in the range   $\Lx\approx 10^{-5.44\,...\,-7.35}\Lbol$. 

The {\em ROSAT} telescope performed an all-sky X-ray survey 
\cite[RASS,][]{vogesetal99-1,vogesetal00-1} in 
the 0.2--2.4\,keV energy band and detected many OB-type stars, confirming the
$\Lx\propto\Lbol$ correlation \citep[e.g.,][]{motchetal97-1}.  
\cite{berghoferetal97-1} demonstrated 
that this correlation flattens considerably below $\Lbol\approx 
10^{38}$\,erg\,s$^{-1}$, that is, for mid- and late-type B stars. 
Since then, the $\Lx\propto\Lbol$ relation  was often revisited and 
confirmed by many independent studies of OB-stars in clusters and in the field. 

With the advent of modern X-ray telescopes \xmm\ and \cxo,  with their broad 
spectral response (0.2--12.0\,keV) and high spatial resolution, the 
interest in X-ray properties of O stars was once more renewed.   
\citet{osk2005} verified the \LxLbol\ correlation selectively, distinguishing
between binary and single stars.   A linear regression analysis of a sample of 
spectroscopic binaries showed a correlation $\Lx\approx 10^{-7}\Lbol$. It was 
found that while binary stars are more X-ray luminous  than  single 
ones, the correlations between X-ray and bolometric luminosities are 
similar for both groups.   
Great effort has been made to study massive star populations 
in individual clusters 
\citep[e.g.,][]{Moffat2002,Wolk2006}.  \citet{sanaetal06-1} observed the 
open cluster NGC\,6231 using \xmm. Based on a rather 
small sample of 12 O stars they found a much lower scatter than previous 
studies (\LxLbol$ = -6.91 \pm 0.15$) and showed that  the $\Lx\propto \Lbol$ 
relation is dominated by soft X-ray emission -- the dispersion becomes larger 
for radiation above 2.5\,keV. 

X-ray emission from a large number of O-stars in the Carina Nebula cluster 
was studied by \cite{nazeetal11-1}.  Using \cxo\ observations of 60 O stars 
in this region,  a ratio of $\LxLbol=-7.26\pm 0.21$ was determined. 
The spectral types  were collected from the 
literature as presented in the contemporaneous  \citet{skiff14-1} catalog. 
Interestingly, using \xmm\ observations of the same region, 
\cite{Antokhin2008} determined $\LxLbol=-6.58\pm 0.79$.  \cite{nazeetal11-1} 
 explained the discrepancy 
 by the different reddening laws and 
bolometric luminosities assigned to O-stars in these studies.

An X-ray catalog of OB-stars  was presented by \cite{naze09-1}. They  
cross-correlated  the  {\em XMM-Newton Serendipitous 
Source Catalogue: 2XMMi-DR3} \citep{watsonetal09-1} and the all-sky catalog of Galactic OB 
stars, which contains $\sim 16000$ known or ``reasonably 
suspected'' OB stars \citep{Reed2003}.
Approximately 300 OB stars were found to have an X-ray counterpart.  
Confirming previous studies, \citet{naze09-1} showed that  X-ray fluxes of O 
stars are well correlated with their bolometric fluxes. However, the   scatter was 
found to be comparable to that of the RASS studies, that is,\ much larger than  
in individual clusters. The average ratio between X-ray (in the 0.5--10.0\,keV 
band) and bolometric fluxes for O-stars was found to be $\LxLbol=-6.45\pm 0.51$.
Although the  \cite{Reed2003} catalog of OB stars  is a valuable resource, it 
is heterogeneous by nature, harvesting data from the SIMBAD 
data-base,\footnote{\cite{wenger00-1}} and might contain dubious  spectral 
classifications.  

While significant effort firmly established an $\Lx\propto\Lbol$ 
 correlation for O stars, its origin is still unknown. A  standard  
explanation of  X-ray emission from O stars is the presence of 
plasma heated by shocks intrinsic to radiatively driven stellar winds 
\citep{lucy+white80-1, 
Owocki1988, Feld1997}. The X-ray luminosity is therefore expected to 
correlate with stellar wind parameters.  One of the most  detailed and careful 
analyses  to date was performed by \cite{sciortino90-1}, who investigated 
correlations between X-ray luminosity and stellar and wind parameters based on a 
large sample of O stars detected by the {\em Einstein} observatory. 
They found that \Lx\ correlates well with mass, the Eddington factor 
(that describes relative influence of radiation pressure with respect to 
gravity), \Lbol, wind momentum (\Mdot\Vinf), and kinetic wind luminosity 
(\Lkin),  but only  weakly with mass-loss rate (\Mdot) and with terminal wind 
velocity (\Vinf). Moreover,  \cite{sciortino90-1} found that none of these 
parameters alone can account for the observed dispersion in  \Lx,  but that 
a combination of  \Lbol, \Vinf\, and \Mdot\ is needed.    Motivated by the 
well-known correlation of \Lx\ with rotation rate in solar-type stars, 
\citet{sciortino90-1} searched for similar correlations in hot stars but could 
not find any evidence. 

Alternative and still speculative explanations for X-ray emission from O stars 
invoke stellar magnetism. In this case, hot X-ray-emitting plasma could 
be associated with stellar spots  \citep[e.g.,][]{WC2007}. The 
presence of such magnetic structures in normal stars is gaining support 
both theoretically and observationally \citep{Cantiello2011, Ram2014}.   

The aim of the study presented in this 
paper is to build a homogeneous catalog of X-ray emitting O stars and to study 
the dependence of their X-ray emission with stellar and wind properties. 
The time is now ripe for such a study.  The Galactic O-Star Spectroscopic Survey 
(GOSS) by \cite{sotaetal11-1, sotaetal2014} and \cite{maiz-apellanizetal16-1} 
is now available.  GOSS consists of more than 590 O stars with 
 spectral types  homogeneously determined from optical
spectroscopy. It is complete down to magnitude 
$B=8$\,mag. As noted by \cite{sotaetal2014},  the  previous studies were 
subject 
to discrepancies in spectral type determinations which can lead to errors
deriving parameters that depend on spectral types (and propagate 
to X-ray studies). The X-ray catalog we present in this paper is  limited to 
stars  from the GOSS  sample for the sake of consistency. 

The release  of  the GOSS  coincides with  an on-going improvement and 
expansion of  {\em The XMM-Newton Serendipitous Source Catalogue}. 
In this paper we consider  GOSS  stars with X-ray counterparts in 
the latest \xmm\  catalog at the time of writing.  We   do not  include the
information from other X-ray surveys and telescopes in order  to  
study the homogeneous  X-ray sample representative for the O star  
population  within our Galaxy. The catalog we present here 
incorporates distances based on the  parallaxes included in the Gaia 
second data release (DR2).

The paper is structured as follows: in Sect\,\ref{sec:asso} we describe the
X-ray, optical, and infrared associations.  Determinations of stellar and X-ray 
parameters are described in Sects.\,\ref{sec:parameters} and \ref{sec:xray}, 
respectively. Correlations between X-ray and wind parameters are discussed in 
Sect.\,\ref{sec:windparams},
and conclusions are drawn in Sect.\,\ref{sec:concl}.

\section{The X-ray, optical, and infrared associations of GOSS and the 
resulting catalog}
\label{sec:asso}

The \xmm\ satellite has been operating since 2000 \citep{jansenetal01-1}. There 
are five X-ray  instruments on board: the three European Photon Imaging Cameras 
(EPIC), and two  Reflection Grating Spectrometers (RGS)
\citep{struderetal01-1,turneretal01-1,denherderetal01-1}. Among
the three EPIC cameras, one is equipped with a pn detector and two with Metal Oxide Semi-conductor CCD arrays (MOS cameras).
These cover the energy range from 0.2\, to 12\,keV,  with an energy
resolution  $E/\Delta E\approx 50$ at 6.5\,keV and with a 
position accuracy
better than 3\arcsec\ (90\% confidence radius).  The catalog we present here 
is limited  to X-ray sources detected by the PN camera.  

Thanks to the \xmm\  large field of view and its high sensitivity, in each 
pointed  observation up to 100  sources are detected serendipitously   
\citep{rosenetal16-1}.
Since 2003 the \xmm\ Survey Science Center has been compiling information 
on serendipitously detected X-ray sources and releasing it to the community in the 
form of catalogs.  

The most recent catalog, released in July 2016, is the 3XMM-DR7; it
contains 727\,790 detections for 499\,266 unique sources, covering a sky area
of about 1\,032 square degrees. The median flux 
in the total photon energy band (0.2\,--\,12\,keV) 
is  $\sim 1.9\times10^{-14}$\,\ergcms; 
in the soft energy band (0.2\,--\,2\,keV) the median flux is 
$\sim 6\times10^{-15}$\,\ergcms, 
and in the hard band (2\,--\,12\,keV) it is
$\sim 8\times10^{-15}$\,\ergcms. About 23\% of the sources have total fluxes
below $1\times10^{-14}$\,\ergcms.  

To find X-ray counterparts of GOSS stars, their optical positions were 
cross-matched  with the 3XMM-DR7 catalog. We looked for all 
possible 
X-ray counterparts within a $3.4\sigma$ search radius, where $\sigma$ is the 
combined positional error of
the X-ray sources and the optical position of the O stars: $\sigma =
\sqrt{(\sigma_{\rm X})^2 + (\sigma_{\rm opt})^2}$. Here $\sigma_{\rm X}$ and 
$\sigma_{\rm opt}$ are the X-ray and optical 
positional
errors respectively and $\sigma_{\rm opt}\,=\,1\arcsec$  is assumed. 

Thanks to the good angular resolution of \xmm, the source  confusion is low -- 
there are only a  few cases where two O stars are unresolved in X-rays
(e.g.,\ HD\,37742 and HD\,37743 both have positions compatible with
3XMM\,J054045.5-015633), or where a star has two possible X-ray associations
and we thus cannot decide which one is the true counterpart (e.g.,\ Cyg OB2-22 A 
is
compatible with both 3XMM\,J203308.7+411316 and 3XMM\,J203308.8+411318). 
We visually inspected these cases, but  concluded that 
O star and X-ray source  cannot be associated unambiguously. After 
discarding these objects,  our final sample contains  135 stars from the GOSS  
uniquely associated with  a 3XMM-DR7  X-ray source. 

As a next step we collected the optical and infrared photometry for our sample 
stars. Our sample catalog  was cross-matched with the GSC2.3.2 and the 
2MASS Point Source Catalogs (2MASS-PSC). The former is
an all sky catalog based on the all-sky photographic surveys from the Palomar
and UK Schmidt telescopes (DSS),  
while the latter  is an all sky survey covering approximately the 4 to 16 
magnitude range in three bands ($J, H$ and \Ks) \citep{cutrietal03-1}. 
Except for HD\,93128,  all our sample  stars  have counterparts  within 
1\arcsec\ in the GSC2.3.2 and the 2MASS-PSC catalogs. 

We exclude  HD\,93128 from the follow up study since it is located in a 
crowded field and its photometry is not reliable, as well as the known  
high-mass X-ray binaries HD\,153919 and LM\,Vel  
\citep{liuetal06-1,martinez-nunez17-1}. This reduces our sample 
to 132 O stars detected with the \xmm\ PN camera and uniquely identified.  

The full sample of our stars is presented in a table that is
accessible in electronic form via the VizieR 
database\footnote{\cite{ochsenbeinetal00-1}}. Table\,\ref{t:cat} 
shows an example of the parameters provided for each star in our 
catalog, which includes distances,  X-ray, and stellar properties. 

\input{table_catalog.tex}

\section{Stellar parameters}
\label{sec:parameters}

\begin{figure}[t!]
\includegraphics[width=\linewidth]{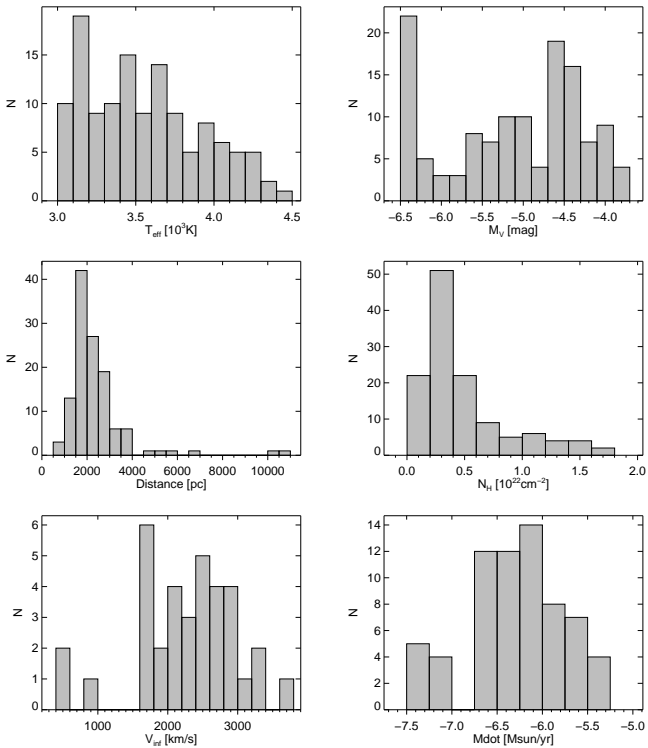}
\caption{Histograms showing distributions of stellar parameters for O-stars in 
our catalog. From left to right and from top to bottom: effective temperatures, 
absolute magnitudes, parallactic  distances, neutral hydrogen column 
densities, stellar wind velocities, and mass-loss rates
for the whole sample (127 O stars shown in light gray), and for stars with
  known terminal velocities (35 stars shown in dark gray).}
\label{g:params}
\end{figure}

All our sample stars have well determined spectral types. This provides 
the basis for  assigning fundamental  parameters to each object. To 
estimate stellar wind parameters we use an empiric approach.  
We collect stellar wind parameters from the most recent literature for a 
sub-sample of our objects, and extrapolate them to the rest of our sample 
stars. Figure\,\ref{g:params}  shows the distribution of the  effective 
temperatures, absolute magnitudes, distances, neutral hydrogen column 
densities, stellar wind velocities, and mass-loss rates for our sample stars.

\subsection{Fundamental stellar parameters}
\label{sec:params}

Stellar effective temperatures, \logg, masses, radii and absolute 
magnitudes are assigned to each sample star according to its spectral type 
using the calibrations from \cite{martinsetal05-1}.
Since in this work the tabulated values are limited 
to luminosity classes V, III, and I,  the values  for luminosity classes IV 
and II were interpolated. For luminosity classes Ia, Ib, and Iab, we assumed 
absolute magnitudes and effective temperatures of supergiant stars.
The peak in the absolute magnitude distribution seen in Fig.\,\ref{g:params} at 
about $-6.4$\,mag corresponds to O supergiant stars. 
To estimate errors on stellar parameters we assumed an uncertainty on the spectral
  subtype of $\pm1$, we then determined the corresponding stellar parameters and
  fixed errors to the maximum difference between these determined stellar parameters; for example, \hbox{\small{$\sigma_{\Teff_{,SpT}} = \max(|\Teff_{,SpT}-\Teff_{,SpT-1}|,|\Teff_{,SpT}-\Teff_{,SpT+1}|)$}}.
  For estimating $\sigma_{\logg}$ we performed a similar analysis but assuming an uncertainty
  on the luminosity class of one subclass. 

Among the sample of O stars studied here, 38 sources are known to be 
spectroscopic binaries \citep{sotaetal2014}. For those cases we display 
the spectral type of the primary stars. Three sources are known to be magnetic: 
$\theta^1$Ori,  HD\,57682,  and  CPD\,-59\,2629 \citep{Donati2002,Grunhut2009, 
nazeetal14-1}.  

\subsection{Extinction}
\label{sec:reddening}

Extinction was calculated using  color excesses \EBV, $E_{J-K_{\rm s}}$ , and 
$E_{H-K_{\rm s}}$ in the optical and in the infrared.  The intrinsic colors 
are from
\cite{martins+plez06-1}.  
In the optical we calculated the extinction using the relation
\Av\,=$R_{\rm V}\times\,$\EBV\ with  \Rv\,$=\,3.1$. 
In the infrared we used the extinction
relations \Ak\,=$\,0.67\times E_{J-K_{\rm s}}$  and
\Ak\,=$1.82\times E_{H-K_{\rm s}}$  from
\cite{indebetouwetal05-1}. Then, the extinction relation 
\Ak$\,=\,0.114\times$\Av\ \citep{cardellietal89-1} provides an alternative way 
to estimate \Av. 

For a set of O stars in  common with \cite{jenkins09-1}  we compared the 
color excess \EBV\   and found that, in general, the values we derive here 
agree well with \cite{jenkins09-1}. We also compared  the optical excesses 
obtained 
from  the infrared colors. Although we find a good general agreement,  
the scatter is larger than when using optical colors only. Therefore  to 
estimate the extinction for our program stars  we use \EBV\ as derived from 
optical photometry. 

There are various  relations  in the literature that link the neutral hydrogen 
column density to the color excess, which range from  
$\Nh\,\sim\,1\times10^{21}\times \EBV$\,cm$^{-2}$ to
$\Nh\,\sim\,9\times10^{21}\times \EBV$\,cm$^{-2}$
\citep{predehletal95-1,gudennavaretal12-1,liszt14-1}.
In this work we adopt $\Nh = 5.6\times10^{21}\times \EBV$\,cm$^{-2}$ 
\citep{bohlinetal78-1}. In Sect.\,\ref{sec:ecf}  we consider in  
detail how this choice affects the estimates of  X-ray fluxes of our sample 
stars.  

For two objects  ($\upsilon$\,Ori and 15\,Mon\,A) the estimated 
\Nh\ is very low. We set \Nh\ for both stars equal to a minimum value of $10^{18}$cm$^{-2}$.

\subsection{Distance}
\label{sec:distances}

The distances to our sample stars were estimated by using 
both the spectrophotometric and the parallactic methods. 

We started by estimating distances from the absolute 
visual magnitudes derived using bolometric corrections  and  the 
photometry of our program stars.  The $V$ magnitudes 
were corrected for the extinction (see Sec.~\ref{sec:reddening}) and the 
errors on the spectrophotometric distances were derived by propagating 
the errors on magnitudes and extinction.

As a next step, we searched the extended Gaia DR2 catalog \citep{BJ2018}, 
and found that the parallactic distances  are  estimated  for 127 of our sample 
stars. These were compared with the  spectrophotometric  distances we derived.  
Reassuringly, the spectrophotometric distances agree well with the  parallactic 
ones. Therefore, in the following we use the Gaia DR2 parallactic distances.

While most of the GOSS sources are within 4\,kpc, six sources are 
estimated to be at larger distances: CPD -59 2600 (estimated distance
$\emph{d}=4.9$\,kpc) is a member of the open cluster Cl\,Trumpler\,16 for 
which distances in the literature range from 2 to 5.4\,kpc 
\citep{2009MNRAS.400..518M,2010MNRAS.403.1491P}. 
HD\,168076\,AB (estimated distance $\emph{d}=5.1$\,kpc), is a member of the open
cluster M\,16 at a distance of 1.7\,kpc \citep{2005A&A...438.1163K}. HD 159176 
(estimated distance $\emph{d}=6.6$\,kpc) and HD 93403 (estimated distance 
$\emph{d}=10.2$\,kpc) are found at distances of 1 and 2.4\,kpc,\ respectively, according to
\cite{2009A&A...507..833M}. Finally, for LS\,4067\,AB (estimated distance 
$\emph{d}=10.7$\,kpc) and ALS\,18\,769 (estimated distance 
$\emph{d}=5.5$\,kpc), we found no distances in the
literature.

\subsection{Terminal wind velocities and mass-loss rates}
\label{sec:vinf-massrate}

The O star winds are radiatively driven \citep{CAK1975}. While grasping 
the fundamental physics, currently the  stellar wind theory has limited 
predictive power. At present, it is  not yet possible to 
theoretically calibrate stellar wind parameters in dependence on spectral 
types. In this situation, different approaches to estimating wind parameters for 
a large sample of O stars are possible. First,  a full multi-wavelength spectroscopic 
analysis of each star in a sample could yield their stellar and wind properties. 
This method is impractical due to, for example,\  observational challenges. Besides, any 
stellar atmosphere model used for spectral 
analysis has its own limitations. Second, the  fundamental
stellar parameters could be used as input for  theoretical recipes  
predicting  wind velocities and mass-loss rates. This is the 
easiest and therefore most commonly used option --  convenient 
routines  are publicly available \citep{vinketal2000}. In 
Table\,\ref{t:cat} we list the theoretical mass-loss rates 
for each star in our sample ($\log\Mdot_{\rm Vink}$). However, there are notable 
discrepancies between the empirically measured mass-loss rates and those
predicted by the theoretical recipes \citep[e.g.,][]{Full2006}. Especially 
dramatic is the situation for late O-dwarfs, where the models fail  
\citep{ Muijres2012}. Such stars constitute a significant part of our sample. 
Third, a semi-empirical approach that builds on existing measurements 
is a practical way to estimate wind parameters for a large number of stars.  
From  UV spectroscopy,  stellar wind velocities are securely measured for 
35 stars of our sample. Furthermore, recent spectroscopic analyses  
using sophisticated stellar atmosphere models  were performed for 25 stars 
in our sample (see Table\,\ref{t:mass-rates}).  Importantly, these stars 
belong to different spectral types and were analyzed with different independent 
stellar atmosphere models.  Hence, the number of empirically well studied stars 
is sufficiently large to establish an empirical calibration of wind properties 
that show a dependency on spectral type.  We choose this option to study the 
dependence of X-ray luminosity on stellar wind parameters.

Using UV spectra of resonance lines  obtained by the  {\em International 
Ultraviolet Explorer},  \cite{prinjaetal1990} measured the terminal  
velocity in  a large number of OB-type  star winds\footnote{We note that 
the errors on the velocity measurements are not provided by 
\citet{prinjaetal1990}}; among 
them 35 belong to our sample. Their wind velocities are typically 
above $\sim$1000\,\kms. 
Winds of two stars,  HD\,93521 (\Vinf\,$\sim$\,490\,\kms) and
$\theta^1$Ori (\Vinf\,$\sim$\,580\,\kms),  are unusually slow. 
Another object with a comparably slow wind is HD\,152408 (\Vinf\,=\,955\,\kms). 
Interestingly, while $\theta^1$Or is a well-known magnetic star, the 
search for magnetic fields in   HD\,93521 and HD\,152408 did not reveal any 
\citep{Grunhut2017,Schol2017}.

\begin{figure}[t!]
\includegraphics[width=\linewidth]{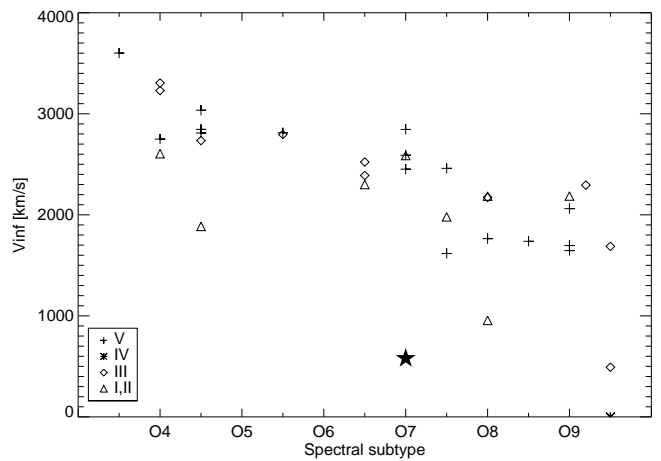}
\caption{Terminal wind velocities (from \cite{prinjaetal1990}) as a function of 
the spectral type for O stars of different luminosity classes (see legend, 
the asterisk denotes a known magnetic star). }
  \label{g:SpT-Vinf}
\end{figure}

As illustrated in Fig.\,\ref{g:SpT-Vinf},  terminal velocity 
decreases with stellar spectral type since stars with lower effective 
temperatures tend to have lower escape velocities \citep{prinjaetal1990}.  
While the stellar wind theory  predicts  a relationship between the 
terminal velocity and the photospheric escape velocity  \citep{Abbott1978, 
LC1999}, the scatter seen in Fig.\,\ref{g:SpT-Vinf} implies 
that the situation  in real objects may be  more complicated, further 
justifying our empirical approach.

Although only 35 stars have measured terminal velocities,
  they cover the whole range of possible stellar parameters (see Fig.~\ref{g:params}),
  and therefore can be taken as representative of the whole sample.

\input{table_mass-loss-rates.tex}

While stellar wind velocities could be directly measured from UV 
spectra, measuring mass-loss rates typically requires the model fitting  
of spectral lines. The errors on the measured mass-loss rates are largely  
systematic, reflecting the limitations of the applied models. Presently, the 
most sophisticated stellar atmosphere models  do not rely on the 
assumption of local-thermodynamical equilibrium (LTE)  (i.e.,\ are non-LTE),
include line-blanketing,  and account for wind clumping 
\citep[e.g.,][]{Hil1998,Hamann2003, Puls2005}. 

We collected literature where the mass-loss rates of our sample stars are 
obtained from spectroscopic analyses done with advanced model atmospheres.  
To the best of our knowledge, we found such analyses for 25 O stars in our sample.  
Each of these models has its own way of describing the effects of wind clumping. 
The mass-loss rates  are  compiled  in  Table\,\ref{t:mass-rates} and  are the 
face values from the corresponding papers (except for luminosity class II-III; 
see discussion below). 

Among the main problems affecting empirical measurements of mass-loss rates
in O stars is stellar wind clumping. Significant effort has been dedicated 
to evaluating the effects of wind  inhomogeneity on spectral analysis and to
improve model stellar atmospheres \citep[see e.g.,][]{Osk2016}.  The current 
understanding is that the in case of   O supergiants,  the theoretical recipes  
\citep{vinketal2000} predict mass-loss rates  that are $\sim 1.5-3$ times 
higher than those obtained from  spectroscopic analyses that  include the 
effects of optically thin (``microclumping'') as well as optically thick 
clumping (``macroclumping'') \citep[e.g.,][]{oskinovaetal07-1, Sund2011, 
Surlan2012, shenaretal15-1}.

\citet{bouretetal12-1} presented  measurements of wind parameters in
a sample of eight Galactic O supergiants. They neglected macroclumping 
(i.e.,\ it was explicitly assumed that all clumps are optically thin).
In order to achieve consistent fits of lines in the optical and UV,
quite small clump volume filling factors were adopted (of about 0.05;
see the original paper for exact definitions and details of the spectral 
analysis). On the other hand, \citet{Surlan2013} used the 3D Monte-Carlo 
radiative  transfer model  to measure mass-loss rates for five Galactic O 
supergiants. The mass-loss rates they derived are systematically higher by
a factor of two compared to those of \citet{bouretetal12-1}. 
\citet{Surlan2013} emphasized that an adequate treatment of the line formation 
in inhomogeneous winds is a prerequisite for interpreting O-star spectra and 
determining mass-loss rates. They showed that mass-loss rates of O-type 
supergiants derived using macroclumping techniques are lower by a factor of 1.3 
to 2.6 than the mass-loss obtained from the theoretical recipes by 
\cite{vinketal2000}. Therefore, 
comparing the results obtained by different models with the results of the  
theoretical recipes, we conclude that the spectroscopically derived  mass-loss 
rates of O-type supergiants are accurate within a factor of approximately three.

\begin{figure}[t!]
\includegraphics[width=\linewidth]{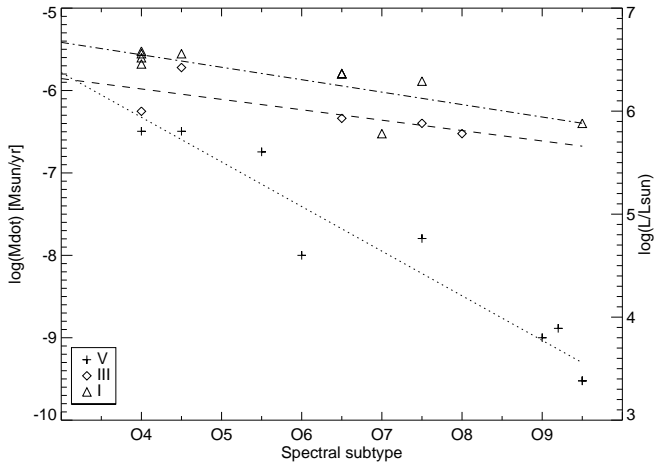}
\caption{Empirically derived mass-loss rates as a function of the spectral
  type for O stars of luminosity classes I, III, and V (see legend). Lines show
 linear fits for each luminosity class. The mass-loss 
rates of giants and supergiants are uncertain within a 
factor of 2-3, i.e., $\Delta\log\Mdot\approx 0.3-0.5$}
\label{g:SpT-Mdot}  
\end{figure}

To our knowledge, only a few Galactic O-type giants have been analyzed by 
advanced clumped stellar atmosphere models 
\citep{Martinsww2005,shenaretal15-1}. On the other hand, a significant number of O-type 
giants were included in the  study of \citet{repolustetal04-1}, who  measured 
mass-loss rates using optical spectra and presented the results without correcting 
for clumping effects. As they note, these values tend to overpredict the real  
mass-loss rates. To correct the mass-loss rates in giant stars, we note that 
\citet{oskinovaetal07-1} have shown that the emission in H$\alpha$ line is not 
affected by macroclumping even in the dense winds of O supergiants,  and therefore 
can be used as a good mass-loss rate indicator. However, the derived mass-loss 
rates still have to be corrected because of optically thin clumping. Adopting a conservative value of 0.1 for 
the volume filling factor,   the
O-giant mass-loss rates from \citet{repolustetal04-1} are reduced by a factor 
of three.  The resulting mass-loss rates are given in Table\,\ref{t:mass-rates} 
and Fig.\,\ref{g:SpT-Mdot}.   

To illustrate the uncertainties associated with the mass-loss measurements and 
justify our approach, it is useful to consider an example: the star\ HD\,93250 is classified as  O3V((f))  in \citet{repolustetal04-1}, who 
obtained its mass-loss rate $\log\Mdot=-5.46^{+0.11}_{-0.16}$. Assuming 
a filling factor of 0.1, the mass-loss rate should be reduced to $\log\Mdot=-5.9$. 
\citet{Martinsww2005} classify HD\,93250 as O3.5V((f+)). Using clumped models 
with a filling factor of 0.01, they  
measure its mass-loss rate as $\log\Mdot=-6.25\pm 0.7$. On the other hand, the 
theoretical mass-loss rate for this star is an order of magnitude higher, 
$\log\Mdot_{\rm Vink}=-5.25$.  In GOSS, the star is re-classified as O4III. Correspondingly, the theoretical mass-loss rate is $\log\Mdot_{\rm Vink}=-5.42$.
This example illustrates the large discrepancy between theoretical and empiric mass-loss 
rates, but it shows that the empirical measurements agree within a factor of  two to three when clumping is taken into account.  This example also illustrates 
that  the errors are largely dominated by the systematic differences between 
the models.  Furthermore,  each atmosphere model has errors on 
mass-loss rates that are due to the errors on the fundamental stellar 
parameters, distance, and the adopted description of clumping.

The mass-loss rates of O-type dwarfs are much less certain compared to 
giants and supergiants. Still, the empirically derived mass-loss rates of 
low-luminosity O dwarfs are orders of magnitude lower than predicted by 
theoretical recipes. This problem is often referred to as ``the 
weak wind problem'' \citep{Martinsww2005}. It is interesting to 
note that nearly all solutions suggested to resolve the weak-wind 
problem in O stars invoke X-rays \citep[][and references therein]{Huen2012}. 
The empirically measured mass-loss rates of well-studied Galactic O-dwarfs 
are given in Table\,\ref{t:mass-rates}. The strong dependence of their 
mass-loss rate on the spectral type is clearly seen  in 
Fig.\,\ref{g:SpT-Mdot} which shows empiric mass-loss rates 
as a function of spectral type for each luminosity class. It is 
interesting to note that the dependence of mass-loss rate on spectral subtype 
is much more pronounced in dwarf stars than in supergiants.

We performed a linear fit to the recent empiric \Mdot\ estimates 
(as described above) and obtained the following scaling relations, 
\begin{equation}
\log(\Mdot [\Msun\,{\rm yr}^{-1}])= -4.2 - 0.5 \times 
\rm{SpT}~~~\rm{for~~OV~stars}, 
\label{eq:dotmv}
\end{equation}
\begin{equation}
\log(\Mdot [\Msun\,{\rm yr}^{-1}]) = -5.5 - 0.1 \times 
\rm{SpT}~~~\rm{for~~OIII~stars},
\label{eq:dotmiii}
\end{equation}
\begin{equation}
\log(\Mdot [\Msun\,{\rm yr}^{-1}]) = -5.0 - 0.2 \times 
\rm{SpT}~~~\rm{for~~OI~stars}, 
\label{eq:dotmi}
\\
\end{equation}
where ``SpT'' denotes a corresponding numeric value of a spectral subtype; for example,\, for an O4I star, the value of SpT is $4$. 

Keeping in mind the significant uncertainties in mass-loss measurements, these 
empiric relations were used to assign the mass-loss rates to the O stars in 
this study. 

\section{X-ray parameters}
\label{sec:xray}
\subsection{X-ray fluxes}
\label{sec:ecf}

The 3XMM-DR7 catalog provides count rates in the 0.2-12\,keV energy band for 
each source. 
To calculate X-ray fluxes on this basis, one needs to establish the energy count 
rate to 
flux conversion factor (ECF). For this purpose we used the Portable, 
Interactive 
Multi-Mission Simulator (PIMMS).  The intrinsic X-ray spectrum 
emitted by an O-type star was approximated by a collisional plasma model, 
{\sc apec} \citep{smith2001},   with solar abundances and single temperature  
$kT= 0.35$\,keV, 
which is typical for single X-ray emitting O stars \citep{nazeetal11-1}. 
For each source in our sample, we calculated  unabsorbed X-ray fluxes from
the mean PN count rate of all detections. The upper and lower values for the 
X-ray 
fluxes based on the mean errors of the count rates were used to estimate  the 
errors of the fluxes. The fluxes were corrected for interstellar absorption 
using 
the \Nh\ values calculated in Sect\,\ref{sec:reddening}.
All the X-ray observations were carried out with medium or thick filters
  meaning that optical loading should not be an issue. Nevertheless, as a sanity check,
  we investigated the summary source detection flag SC\_SUM\_FLAG and confirmed
  that no sources in our catalog were flagged for suspected optical loading.
As  a following step, X-ray luminosities were computed for the distances calculated as 
described in Sect\,\ref{sec:distances}. All stars in our sample have X-ray 
luminosities in the range $10^{31-34}$\,\ergs, for either 
spectrophotometric or parallactic distance estimates.

Among our sample stars, 60 objects have been observed by \xmm\ only once. The 
rest 
of the sources have been observed and detected 3-4 times on average, and up to 
11 times, 
with individual exposure times ranging from 3 to 100\,ks.  The unabsorbed
  fluxes were 
calculated 
using  the average value of the count rates among all detections for each 
source. 
Therefore the fluxes and luminosities presented in this catalog  
characterize  steady X-ray emission of O stars. 

O-type stars show typical X-ray variability on the level 
of  $\sim$10\%\ \citep[e.g.,][]{osk2001, naze2013, Massa2014}. 
Time series were automatically extracted by the \xmm\ pipeline for 108 out of 
the 132 O stars we study here. Seven of the stars are
classified in the 3XMM-DR7 catalog as variable based on a $\chi^2$ variability 
test:  
{$\theta^1$\,Ori C, 
HD\,193322A,
HD\,152218,   
$\sigma$\,Ori\,AB, 
HD\,97434, 
$\zeta$\,Pup,  and 
HD\,93129A}.
An analysis of their X-ray variability  is  beyond
the scope of this paper.    

Besides count rates, the 3XMM-DR7 catalog also  provides X-ray fluxes for all 
objects.    
Comparing them with the fluxes  we derived for our sample stars,  
we find  a mean difference in the flux ratio between the \xmm\ catalog and 
our derived values of $\approx 1.65$. This difference is due to 1) the different 
 adopted X-ray spectral models;  the 
3XMM-DR7 fluxes are calculated under assumption of a power-law type X-ray 
spectrum, while we use a thermal plasma model; and 2) the more accurate
interstellar neutral hydrogen densities towards O stars which we derive in this 
study. 

To place  an upper limit on the X-ray luminosity of an  O-star which was 
observed but not detected by \xmm,  we built  the cumulative distribution
function of X-ray fluxes of all sources detected within the same field of view 
as listed in 3XMM-DR7 catalog.   
The upper flux limit to a non-detected O star could then be calculated as the flux correction factor 1.65
multiplied by the flux at which 90\% of other field sources are detected. 

We searched for non-detections and placed upper limits on their X-ray flux.
  Firstly we searched for O stars within 900\arcsec\ of the center of any
  \xmm\ observation, that is, within the field of view. We ensured that the optical positions
  were within X-ray images, that is, not falling into gaps or outside of X-ray image boundaries
  (especially for observations performed not in the full window model).
  Nine O-type stars were observed but are not detected by \xmm.

O stars not detected in X-rays have late spectral subtype in the range O6--O9.5,
  cover all luminosity classes and have upper X-ray to bolometric flux ratio close to -7.
  Since the number of non-detected stars is small, we do not expect the distribution of
  \LxLbol\ (see Sect.~\ref{sec:lxlbol}) to be biased towards
  high X-ray to bolometric flux ratio.

\begin{figure}[t!]
\includegraphics[width=\linewidth]{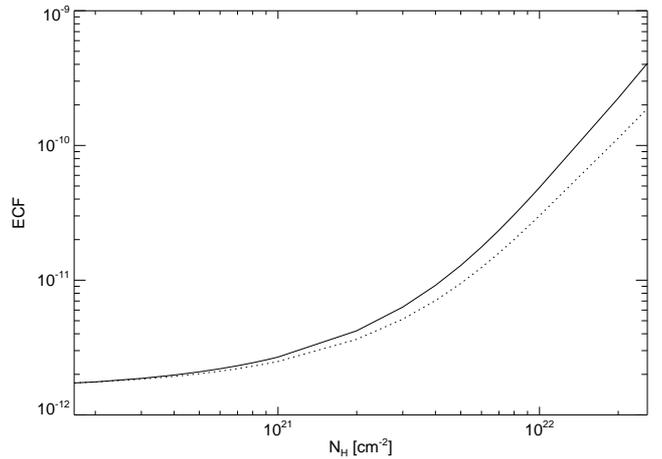}
\caption{ ECF calculated assuming
  an APEC model with solar abundance and temperature of 0.35\,keV (solid line)
  and for 0.5\,keV (dotted line) as a function of \Nh.}
\label{g:ecf} 
\end{figure}

The choice of  the plasma temperature and \Nh\  affects the calculation of 
the ECFs
and, consequently, the estimated X-ray fluxes. To investigate how robust our  
derived X-ray luminosities are,  we calculated the ECFs for two different plasma 
temperatures, $kT = 0.35$\,keV and $kT = 0.5$\,keV, and for a wide range of 
\Nh\ values. As illustrated in Fig.\,\ref{g:ecf},   a different choice of  the
plasma temperature in a plausible range leads to an only slightly different ECF. 
On the other hand, the  ECF's dependence on the \Nh\  increases for 
larger \Nh\ values.   The average  \Nh\ of our sample stars is 
$7\times10^{21}$cm$^{-2}$ (see Fig.\,\ref{g:params}).  As can be seen from 
Fig.\,\ref{g:ecf},  changing this value by a factor 1.5 results in variation of the ECF  by a factor of two. Hence, the ca lculated X-ray fluxes
depend mainly on  the adopted values of \Nh.

There are 34 stars  in common between ours and   \cite{naze09-1} catalogs. The 
fluxes  we obtained for these stars  are a factor of two higher,  chiefly 
because we adopted a different  relation between \EBV\ and  \Nh\  in our 
study.

\subsection{Correlation between X-ray and bolometric luminosity of O-type stars}
\label{sec:lxlbol}

\begin{figure*}[t!]
\begin{center}
  \includegraphics[width=0.75\linewidth]{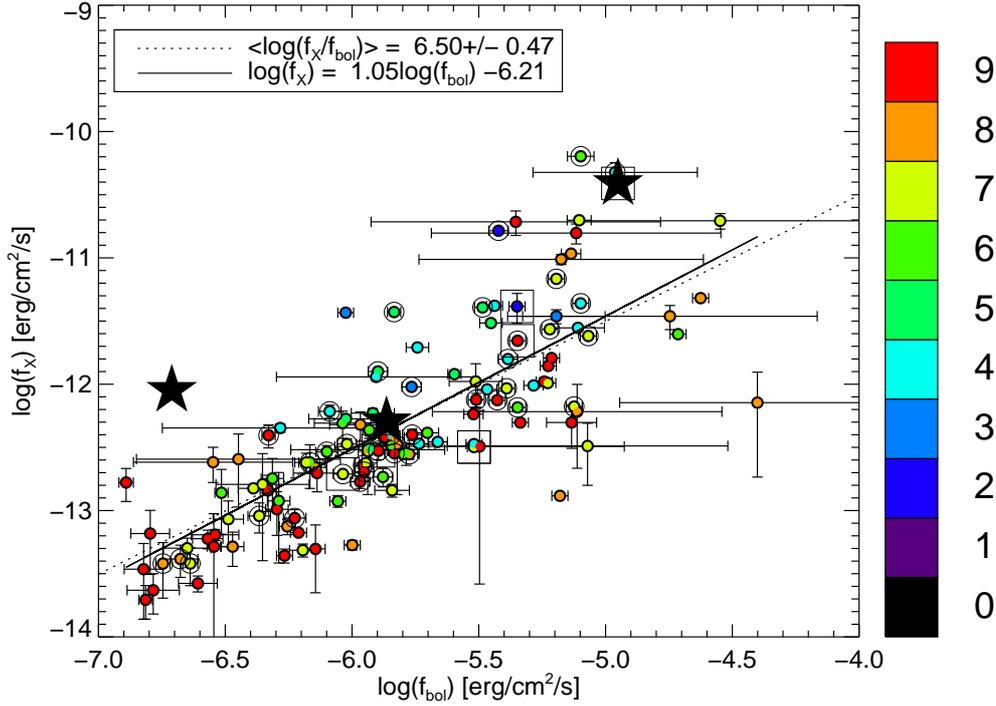}
  \caption{X-ray flux vs. bolometric flux for O stars of all luminosity 
classes and as a
    function of the spectral subtype (color-coded with earlier subtypes being 
bluer, and the later subtypes being redder). Known spectroscopic binaries
    are highlighted with a ring around the solid circle, magnetic O stars
    are represented by asterisks, and stars flagged as variable are
    shown with a squared symbol. The linear fit $\logfx = a + b\times\logfbol$ 
and the mean \fxfbol\
    value are shown as solid and dotted lines, respectively (see text for 
details).}\label{g:fxfbol} 
\end{center}
    \end{figure*}

Considering the relation between X-ray and bolometric fluxes,  \fxfbol, 
allows us to ignore the problems associated with the uncertainties in distances. 
The bolometric fluxes are calculated  according to spectral types of our sample 
stars as 
\begin{equation}
\logfbol = -4.6-(\mathrm{V}-\Av+\mathrm{BC_V})/2.5. 
\label{eq:fbol}
\end{equation}

Figure\,\ref{g:fxfbol} displays the \fxfbol\ relation for the full sample,
independent on their luminosity class. Confirming previous  studies,
we find that the X-ray fluxes scale with the bolometric fluxes. The linear 
regression fit yields the correlation 
\begin{equation}
\logfx = - 6.21 + 1.05\times\logfbol.
\label{eq:fxfbol}
\end{equation}
This result is obtained for the total energy band of the \xmm\ PN camera 
(0.2\,--\,12 keV).
For completeness we include in Appendix~\ref{sec:appendix1} a similar analysis 
carried out
for two different energy bands (a soft and a hard one). 

The bolometric luminosities, \Lbol, were derived in two ways.  The 
distance-independent bolometric luminosities are based on spectral 
type \citep{martinsetal05-1}. The distance-dependent luminosities  ($\Lbol=4\pi 
d^2 \fbol$)  are based on bolometric fluxes as obtained using 
Eq.\,(\ref{eq:fxfbol}) and Gaia DR2 distances (these luminosities are included 
in  Table\,\ref{t:cat}). There is good agreement between bolometric 
luminosities derived by different methods. We use the distance dependent 
luminosity since it eliminates distance dependence in \LxLbol. 

\begin{figure}[t!]
\begin{center}
\includegraphics[width=\linewidth]{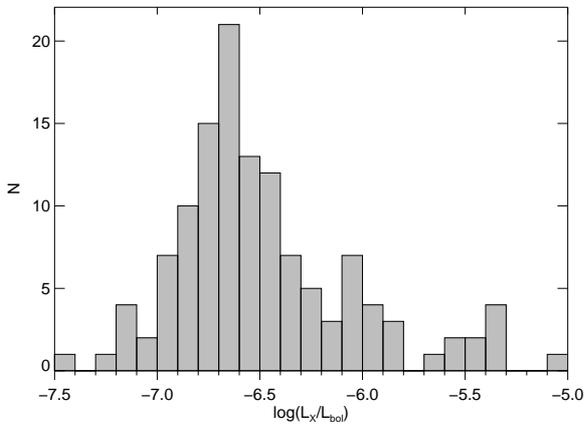}
\caption{Distribution of the X-ray to bolometric luminosity ratio for all the
  stars in the sample.}\label{g:hist_LxLbol_All}
\end{center}  
\end{figure}

Figure\,\ref{g:hist_LxLbol_All} shows the \LxLbol\ distribution for all
our sample stars. The \LxLbol\ distribution is non-Gaussian, with a
peak at around -6.7 and minimum and maximum values at around -7.5 and -5.0 
respectively, corroborating previous results based on ROSAT data 
\citep{berghoferetal97-1}. The impact of errors is discussed in Appendix~\ref{sec:appendix2}. 
  As can be seen from Fig.\,\ref{g:hist_LxLbol_All} the distribution shows a tail towards high values.
  Binarity could result in an increased X-ray flux, for example, due to wind-wind collisions. In the following sections we discuss the dependence of \LxLbol\ on luminosity
  class, spectral type, and binarity.

Although the \LxLbol\ distribution is non-Gaussian, we computed average values for
\fxfbol\ for all stars as a function of the spectral type,
for single and binary stars and for two different ranges of spectral types. The 
results are presented in Table\,\ref{t:meanvalues}.

\input{table_fits.tex}

\subsubsection{The dependence of \LxLbol\  on luminosity class}
\begin{figure*}[t!]
\begin{center}
\includegraphics[width=\linewidth]{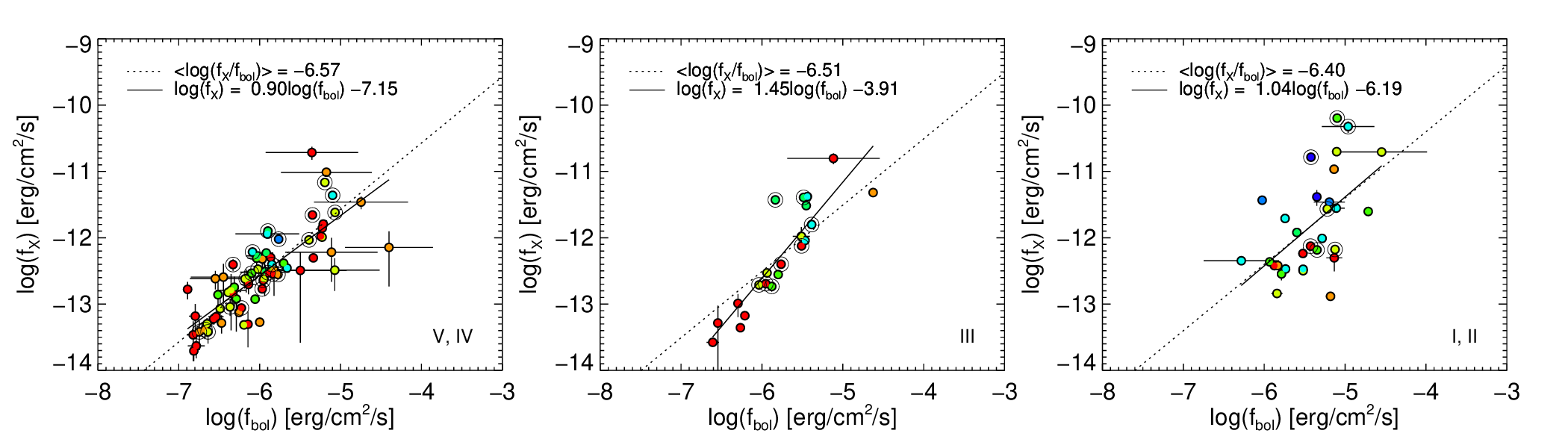} 

\includegraphics[width=\linewidth]{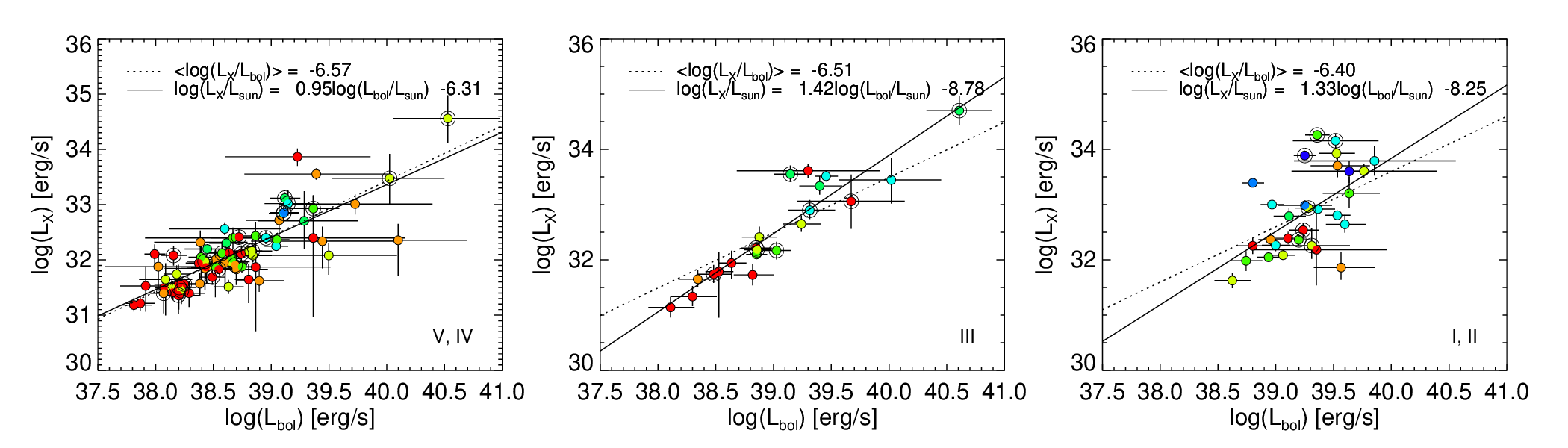}
  \caption{\fxfbol\ (top) and \LxLbol\ (bottom) relation for O stars of 
different luminosity
    classes (V, III, and I from left to right). Known spectroscopic binaries
    are highlighted with a ring around the solid circle, and spectral types are
    color coded as in Fig.\,\ref{g:fxfbol}.
    The linear fits $\logfx = a + b\times\logfbol$ and the mean \fxfbol\
    values for each case are shown as solid and dotted lines, respectively (see 
text for details). For O supergiants, the correlations are weak, while 
they are significant for O dwarf and giant stars (see Table~\ref{t:fitparams}). 
  }\label{g:fxfbol_LCl} 
\end{center}
\end{figure*}

\input{table_stats.tex}

The \fxfbol\ relation for different luminosity classes is shown 
in Fig.\,\ref{g:fxfbol_LCl}. The correlation coefficients between X-ray and 
bolometric fluxes  and the 1-$\sigma$ confidence intervals of the coefficients of the 
linear fits are given in Table\,\ref{t:fitparams}. To measure the strength of 
the linear correlation, we calculated the Pearson correlation coefficient $R$. 
This coefficient can take values between 1 and -1. Values close to 1 or -1 
reflect a strong positive or negative linear correlation between the two variables, 
while values close to 0 indicate no relationship. 

Dwarfs and giant O stars show a very strong correlation between \logfx\ 
and \logfbol. The O6\,IV star ALS\,15108  has unusually high \logfx\ and 
\logfbol. We investigated this star in more detail, and found that it has 
a very high reddening ($\EBV \sim 3.1$). The star belongs to the Cygnus OB2 
association. The extinction is known to be large in the area. It is likely that 
we overestimate the neutral hydrogen column density towards this star, 
leading to its unusually high X-ray luminosity.  

The correlation between the X-ray and the bolometric  
for O supergiants is only moderate\footnote{Interestingly, the weak or absent 
correlation between X-ray and bolometric luminosity was previously noticed in 
another class of massive stars with dense and fast winds, namely Wolf-Rayet 
stars \citep{ignace2000}.}.
Despite the relatively small number of O-supergiants in our study, one can 
see
that confirmed spectroscopic binary supergiants  have approximately constant
bolometric flux ($\logfbol\,\sim\,-5.5$), while  their X-ray fluxes 
span over three orders of magnitude (Fig.\,\ref{g:fxfbol_LCl}).   

\begin{figure}[t!]
\includegraphics[width=\linewidth]{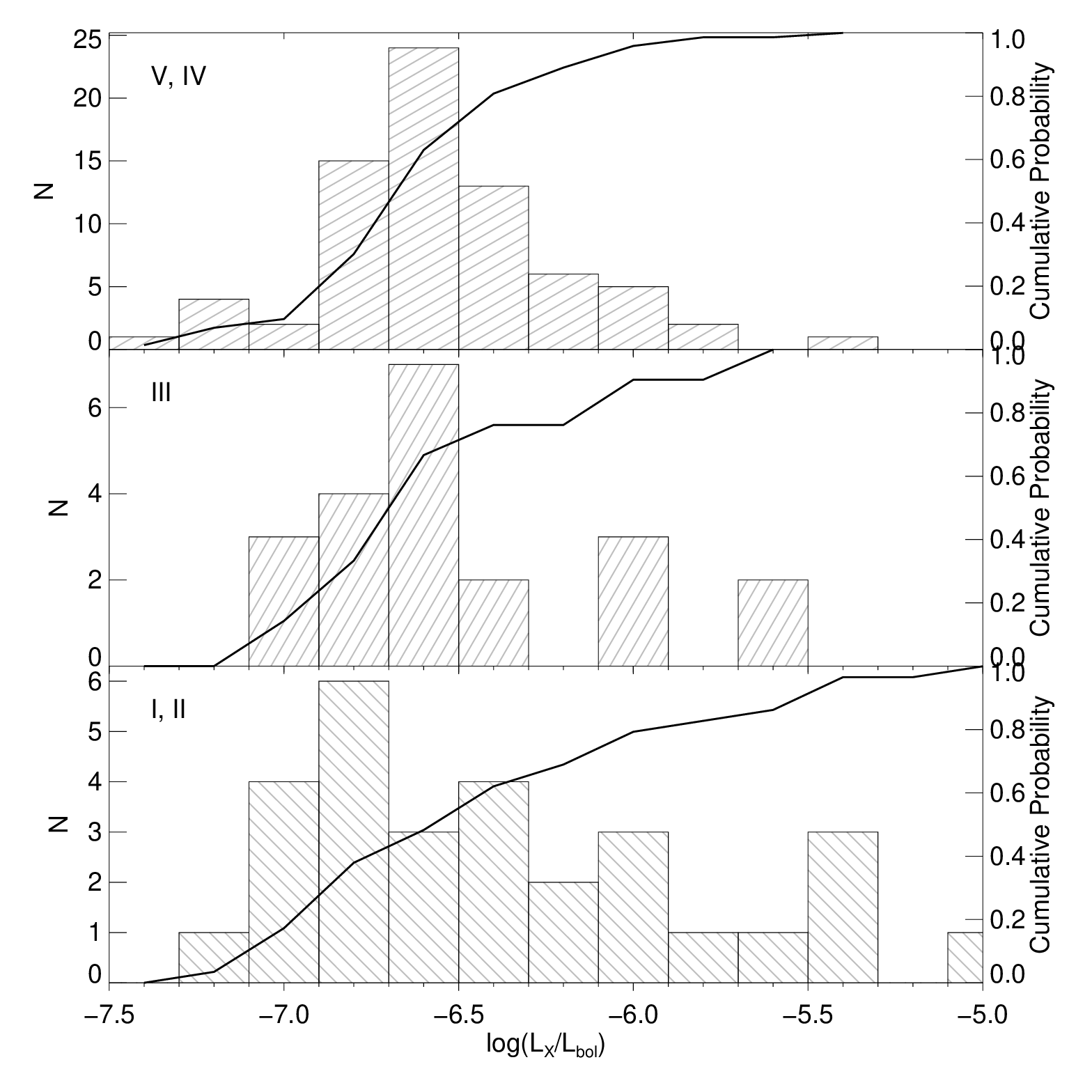}
\caption{\LxLbol\ histogram and cumulative distribution function
  for dwarfs (top), giants (middle), and supergiants (bottom).}
  \label{fig:hist_LxLbol_LC} 
\end{figure}

The \LxLbol\ distribution depends on the  luminosity class, as illustrated in 
Fig.\,\ref{fig:hist_LxLbol_LC}. For dwarf and giant stars the distribution 
peaks at around $-6.6$, shifting towards smaller values, $-6.8$, for supergiant 
stars.  Mean values are $-6.6\pm0.4$, $-6.5\pm0.4,$ and $-6.4\pm0.6$ for 
dwarf, giant, and supergiant stars, respectively. Interestingly, all these 
averages are above the canonical value of $-7$. It is important to keep in 
mind that the \LxLbol\  distribution is non-Gaussian. 

To study the dependence of the \LxLbol\ distribution on spectral type,  we 
divided our sample into two groups -- early O stars with spectral subtypes in 
the range O2\,--\,O6, and late O stars  with spectral subtypes from O6 to O9.5. 
The \LxLbol\ distribution peaks around $-6.4$ for early O-type stars, and 
shifts towards lower values close to $-6.6$ for later spectral types. The mean 
values are $-6.3\pm0.5$ and $-6.6\pm0.4$ for early and late spectral types,
respectively. 

\begin{figure}[t!]
\includegraphics[width=\linewidth]{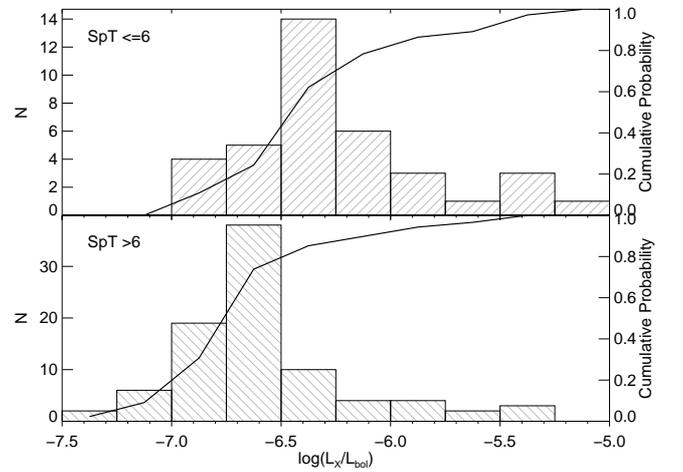}
\caption{\LxLbol\ histogram and cumulative distribution function for 
early  O2\,--\,O6 subtypes (top) and late O6\,--\,O9.5  subtypes
  (bottom).}
 \label{fig:hist_LxLbol_SpT} 
\end{figure}

\subsubsection{\LxLbol\ dependence on binarity}

\begin{figure}[t!]
\includegraphics[width=\linewidth]{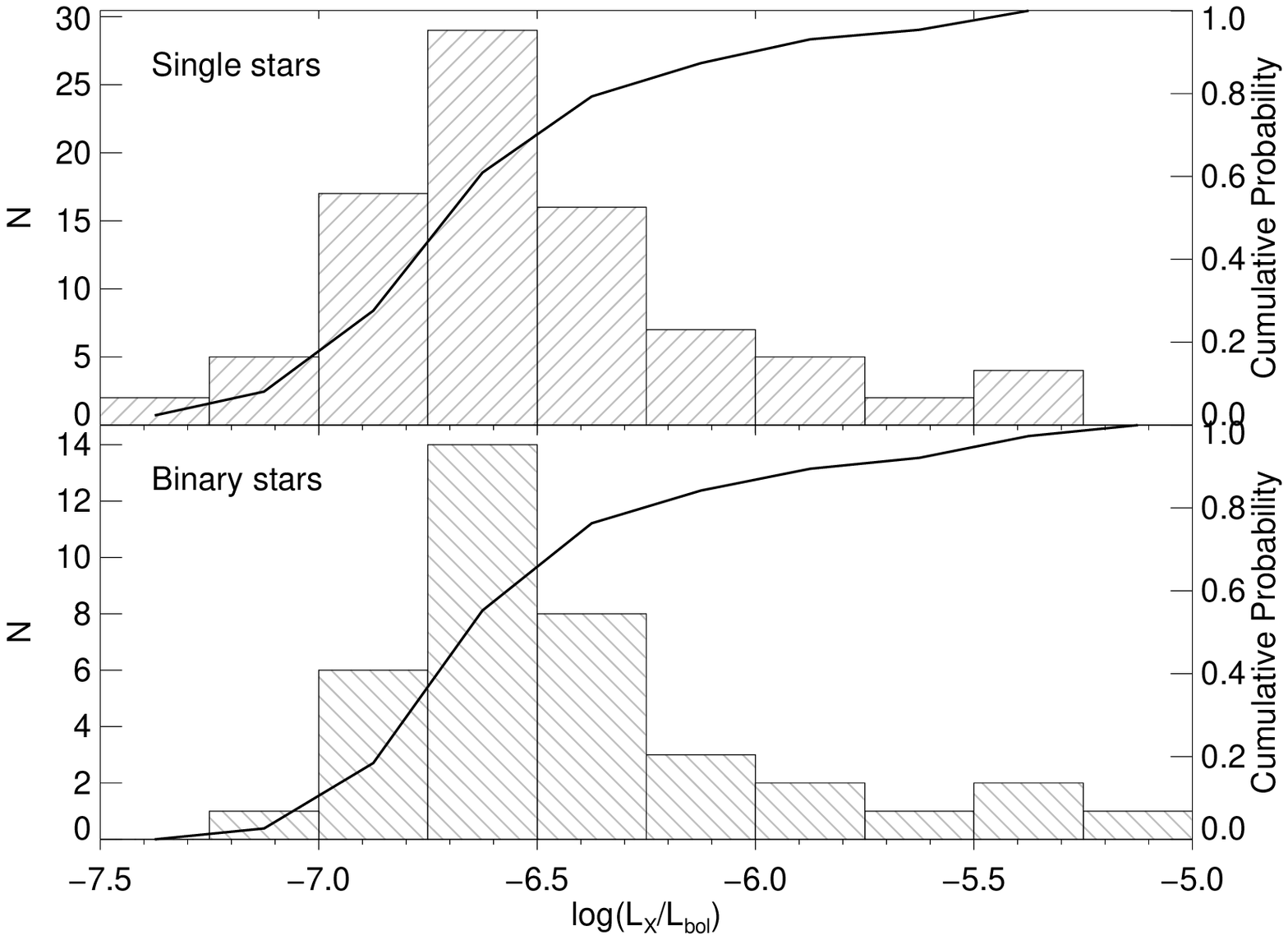}
\caption{\LxLbol\ histogram and cumulative distribution function for 
 putative single stars (top), and  in binaries (bottom).}
\label{fig:hist_LxLbol_bin} 
\end{figure}

The distribution of \LxLbol\ for confirmed binaries is very similar to that  of 
single stars (see Fig.\,\ref{fig:hist_LxLbol_bin}). Both distributions peak at
$\LxLbol\,\sim\,-6.6$ and the mean values are $-6.6\pm0.5$ and
$-6.4\pm0.5$ for single and binary stars, respectively. This corroborates the 
results of previous studies   \citep{sciortino90-1, osk2005, naze09-1}.
Most likely, the true fraction of binaries in our sample is significantly 
higher than those of the  already confirmed binaries, and the list of punitively
single  stars is very likely contaminated by binaries. A thorough analysis 
of the confirmed single star sample and confirmed colliding wind binaries  is 
beyond the scope of this paper. 

\subsection{X-ray hardness-ratios}

\begin{figure*}[t!]
\includegraphics[width=\linewidth]{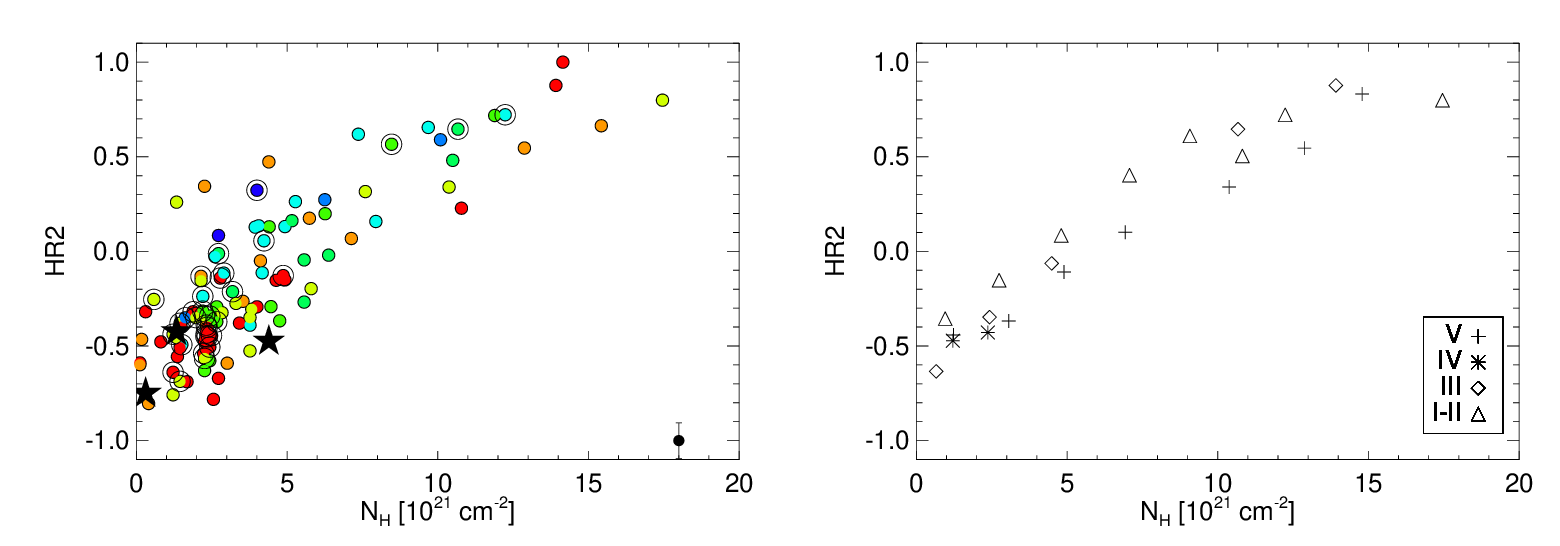}
\includegraphics[width=\linewidth]{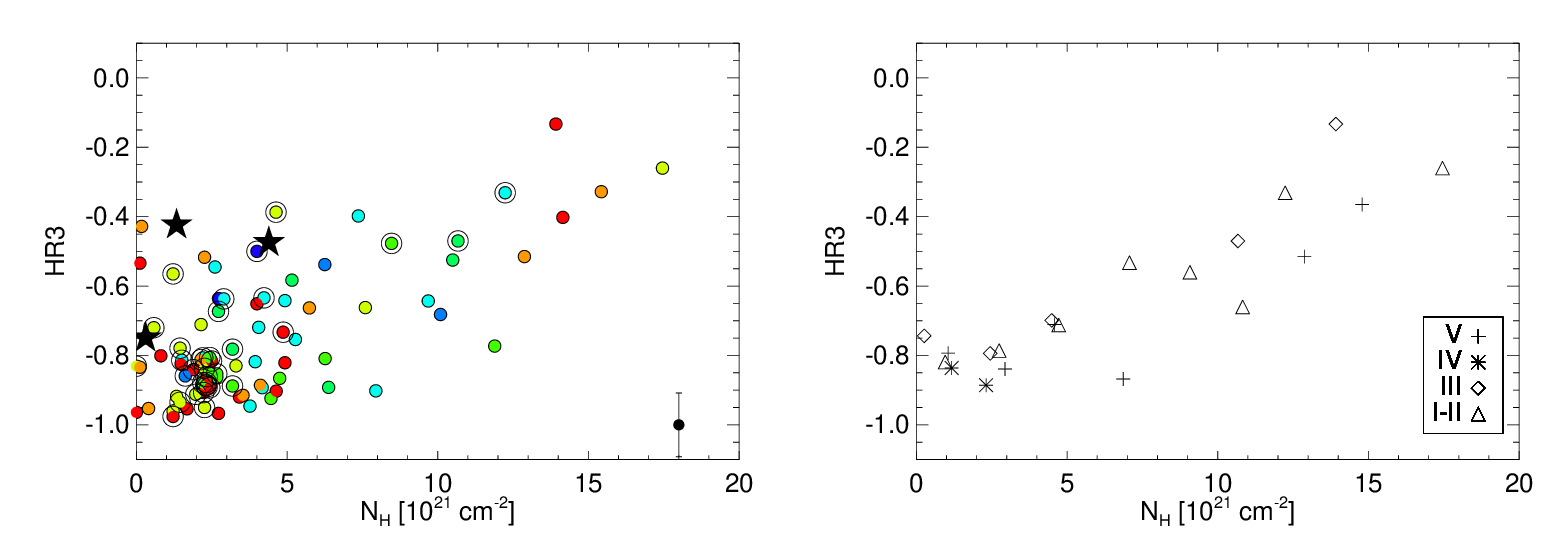}
\caption{Hardness ratios HR2 (top) and HR3 (bottom) as a function of \Nh. In 
the left panels symbols have
  been colored according to the spectral type as in Fig.\ref{g:fxfbol} and mean
  error in the HR is indicated in the lower right corner of the plot. In the
  right panels, mean HRs have been calculated per bin of \Nh\ and different
  luminosity classes; different symbols stand for different luminosity classes
  as indicated in the legend.}\label{g:hr2} 
\end{figure*}

Among other parameters, the 3XMM-DR7 catalog provides hardness ratios (HR)  
describing  the difference in count rates between two consecutive bands, 
defined as 
\begin{equation}
  \mathrm{HR} =
 \frac{\mathrm{count\_rate_{i+1}-count\_rate_{i}}}{\mathrm{count\_rate_{i}
+count\_rate_{i+1}}}. 
\end{equation}
In other words, the HR gives a rough idea of the slope of the X-ray 
spectrum; for example,\ HR\,$=\,0$ would be a flat spectrum, HR\,$>\,0$ implies the 
source emits more photons in the harder band, and HR\,$<\,0$  tells us the 
source emits more photons in the softer band. The HR value $1$ or $-1$ in 
3XMM-DR7  means the source was not detected in either band.

We calculate the mean PN hardness ratios HR2 and HR3 that describe emission 
in the energy bands 2 and 3, and 3 and 4, and cover the 
energy ranges $0.5$\,--$\,1.0$, $1.0$\,-\,$2.0$, and $2.0$\,--\,$4.5$\,keV,
respectively.  

Figure\,\ref{g:hr2} shows the relation between HRs and \Nh\ limited  to sources 
with errors on HRs lower than $0.3$. One can see that, as expected, HR2 
increases with \Nh. There are only a few sources with high HR2 and HR3 at 
a relatively low \Nh, these are intrinsically hard X-ray sources. 
We computed mean HR in bins of similar \Nh\ for different luminosity classes, 
and found 
that supergiant stars have higher HRs at the same \Nh, as can be seen in particular 
in the
HR2 (right upper panel of Fig.\,\ref{g:hr2}) but also in the HR3.
This shows that supergiant stars have systematically harder spectra than 
dwarfs. 

\section{Correlations between X-ray and wind parameters}
\label{sec:windparams}
\begin{figure*}[t!]
\includegraphics[width=0.9\linewidth]{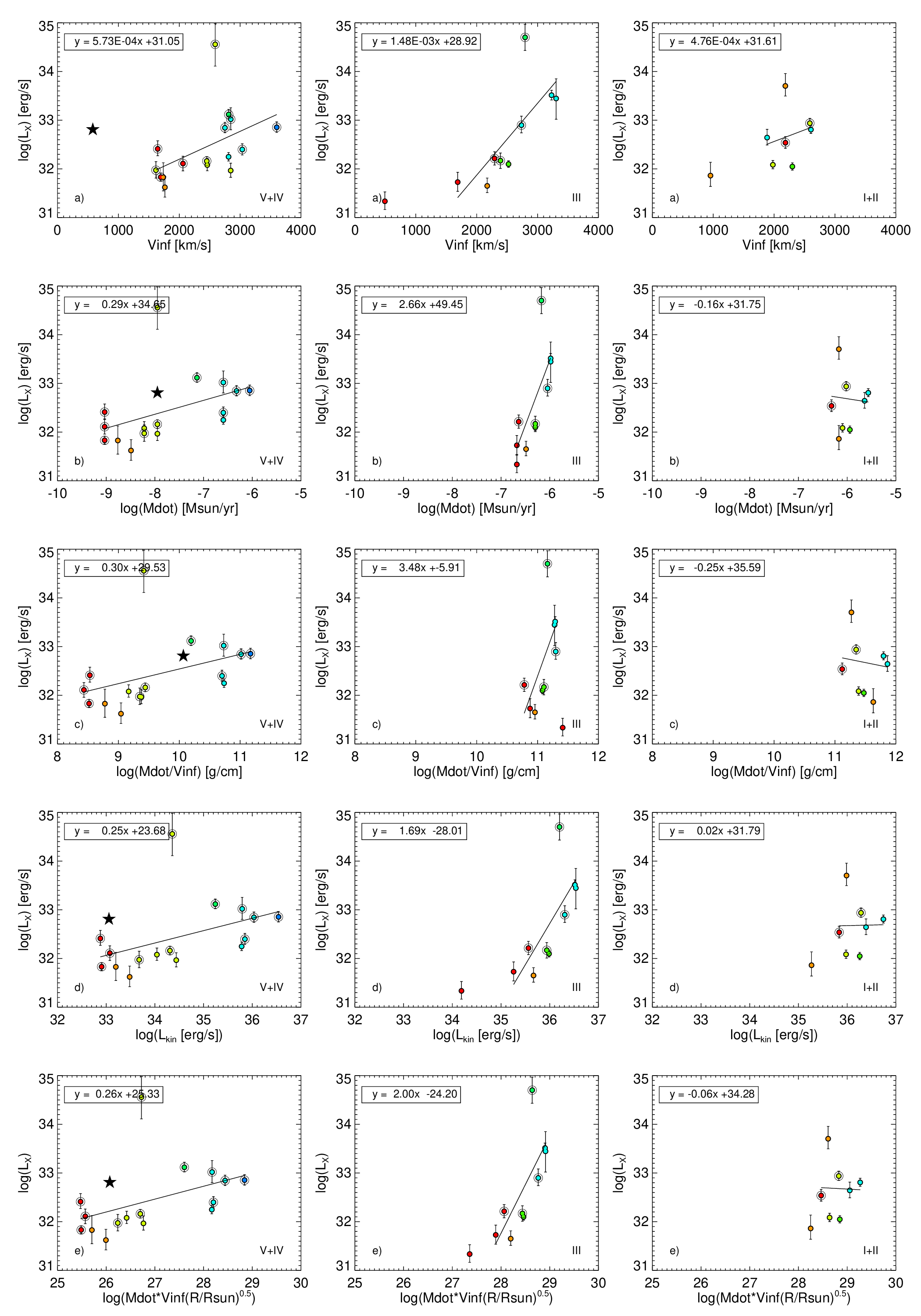}
\caption{X-ray luminosity as a function of wind parameters.  Only 
the stars with wind velocities measured from the UV spectra are shown.  
The mass-loss rates are uncertain within a factor of 2-3 due to systematic errors, 
i.e., $\Delta\log\Mdot\approx 0.3-0.5$ (see Sect.   \ref{sec:vinf-massrate} for a discussion on errors).  
From the left to the right column the results for dwarfs, giants, and 
supergiants are shown.  
The {\it first row from the top} shows the correlations with  wind terminal 
velocity; 
the {\it second row from the top} shows the dependence on mass-loss rate 
calculated using the empirical SpT\,--\,\Mdot\ relations derived in
Sect.~\ref{sec:vinf-massrate}; 
the {\it middle row} shows the dependence on  wind density; 
the {\it fourth row from the top} shows the dependence on wind kinetic
luminosity;
the {\it bottom row} shows the dependence of X-ray luminosity 
on modified stellar wind momentum. The color coding of 
different spectral subtypes is the same as in Fig.\ref{g:fxfbol}. 
Linear fits are shown with a line.}
\label{fig:Ekin} 
\end{figure*}

In this section we investigate the correlations between  X-ray luminosity and
stellar wind parameters for different spectral types and luminosity classes. 
As explained in detail in Sect. \ref{sec:vinf-massrate}, the 
stellar wind velocities were measured from the UV spectra only for 
$35$ stars in our sample. The wind velocities show a general trend of 
decreasing for later spectral types (see Fig.\,\ref{g:SpT-Vinf}), however with 
a significant scatter.   The scatter is also obvious in 
Fig.\,\ref{g:SpT-Mdot} which shows the dependence of mass-loss rates on 
spectral subtype for stars of different luminosity classes.  The uncertainty on 
empirically derived  mass-loss rates is about a factor of three. The errors 
on wind velocities and mass-loss rates propagate to other wind parameters, 
such as wind kinetic energy and momentum. 
  
In Fig.\,\ref{fig:Ekin} the correlation between X-ray luminosity and wind 
parameters is shown. To reduce the errors, only the stars with wind 
velocities empirically measured from the UV spectra are included in this 
figure.   Hence the upper row in Fig.\,\ref{fig:Ekin} displays  only the 
measured quantities of the X-ray luminosities and wind velocities for a 
sub-sample of O stars. 

The linear fits were performed  without taking into account the magnetic stars, 
 and  also omitting HD\,93521, given its unusually slow wind. We 
note that there are three other peculiar systems with rather low \Lx\, and low 
stellar wind density: HD\,14633 (ON8.5V),  BN Gem (O8:V,  the lowest X-ray 
luminosity among the whole sample stars), and HD\,15137 (O9.5II-III). Results 
of the linear fits are given in Table\,\ref{t:lx-fits}; the errors in the 
coefficients correspond to the 1-$\sigma$ confidence interval.

  The correlation between X-ray luminosity and wind velocity is seen for 
dwarf and giant stars, but is absent  for supergiants.  As can 
be seen in  Fig.\,\ref{fig:Ekin}, for dwarf and giant 
stars the X-ray luminosities increase as winds increase in speed. Despite  
relatively large scatter (in particular for dwarfs) which is largely induced by 
binaries with their higher X-ray luminosities, the  overall correlation is weak 
for dwarfs and strong for giant stars. For single stars this correlation 
implies that stars with later spectral types, that is,\ stars with lower effective 
temperatures and lower \Vinf, have lower X-ray luminosities. The situation is 
different for supergiants, where only very weak or no correlation between X-ray 
luminosity and terminal velocity is found. This may reflect the reality or be 
simply due to the small number of supergiants with measured \Vinf.

Since the wind velocities and X-ray luminosities are empiric measurements, 
the detected correlations (or their absence) are physical and not induced by 
systematic errors. This gives us confidence to search for  
correlations between X-ray luminosities and other wind parameters.

Since spectral types are known for all stars in our 
sample we can estimate their mass-loss rates  using the  SpT\,--\,\Mdot\ 
scaling relations Eqs.\,(\ref{eq:dotmv}), 
(\ref{eq:dotmiii}), (\ref{eq:dotmi}) derived in  Sect\,\ref{sec:vinf-massrate}. 
As can be seen from Fig.\,\ref{fig:Ekin} and Table\,\ref{t:lx-fits}, the X-ray 
luminosities show weak correlation with  mass-loss rate for main sequence 
and giant stars, while there is basically no correlation for supergiants.

The correlation is also present when considering spectral types within each 
luminosity class -- later spectral subtypes tend to have lower mass-loss rates 
and lower X-ray luminosities than early spectral subtypes. 

As a next step we investigate the scaling of X-ray luminosity with global 
parameters describing stellar winds, such as the wind kinetic luminosity 
$L_{\rm 
kin}=\,\frac{1}{2}\,\Mdot \times \Vinf^2$ [erg\,s$^{-1}$], the density estimator
$N_{\rm l}\,=\,\frac{\Mdot}{\Vinf}$ [g\,cm$^{-1}$], and the 
modified stellar wind momentum, $\mathrm{D\,=\,\Mdot \times \Vinf \times 
\sqrt{\frac{R_\ast}{R_\odot}}}$, defined according to \citet{kudritzkietal00-1}.
Reflecting the scaling of X-ray luminosity with mass-loss rate and wind 
velocity, the giant O stars show strong correlations with wind 
parameters, while dwarfs show only weak correlations and no correlation is seen
in O supergiants.

The X-ray luminosity of late O-type dwarfs (exhibiting the weak wind 
problem; see Sect.\,\ref{sec:vinf-massrate}) is a few percent of the 
stellar wind kinetic luminosity  (see Fig.\,\ref{fig:Ekin}). This fraction 
drops for earlier spectral subtypes and for giants and supergiants; 
for example,\ in the latter case  $L_{\rm X} < 10^{-3} L_{\rm kin}$. The trend showing 
larger X-ray output from O stars with weaker winds has already   been reported 
\citep[e.g.,][]{oskinovaetal06-1}.  

The density parameter $N_{\rm l}$ is a proxy for the mean wind density, 
and determines the shock cooling length \citep[see Eq.\,(14) and its 
discussion in][]{Hil1993}.  \citet{owockietal13-1} noticed that for radiative 
shocks one can expect that  $\Lx\,\propto\,(\Mdot/\Vinf)^{1-m}$. Using 
the theoretically expected scaling between mass-loss rates and bolometric 
luminosities,  they suggested that if the exponent $m \approx 0.4$ 
the observed scaling between $L_{\rm X}$ and $L_{\rm bol}$ can be reproduced. 

We test this suggestion using our empiric approach. As can be seen from 
Fig.\,\ref{fig:Ekin} and Table\,\ref{t:lx-fits}, for the total ensemble of 
O stars, the correlation between \logLx\ and $\log{N_{\rm l}}$ is very 
weak. Only giants show a significant correlation which is different from 
the one expected theoretically. 

Considering the modified wind momentum, we notice similar trends as for 
other wind parameters -- a significant correlation is observed for dwarfs and 
giants, while no correlation is seen for supergiant stars  
in Fig.\,\ref{fig:Ekin}.
The analysis by \citet{Mok2007} confirmed theoretical expectations that modified 
wind momenta do not 
depend on stellar luminosity class (when not accounting for the weak wind star 
problem). In this study we choose the empiric approach, and for low-luminosity 
dwarfs we adopt empiric mass-loss rates which are significantly 
lower that those predicted theoretically.  Figure\,\ref{fig:Ekin} shows that 
the scaling relations between X-ray luminosity and wind parameters are quite 
different for different luminosity classes. 

To measure the strength of the correlation between X-ray luminosity and wind parameters 
  we calculated the Pearson correlation coefficients. We also used a non-parametric method,
  in particular we calculated the Spearman's rank correlation coefficient $\rho$.
  The Spearman coefficient can take values between 1 and -1. Values close to 1
  or -1 reflect a strong positive or negative monotonic correlation (not necessarily a linear correlation
  such as for the Pearson coefficient) between the two variables, while values close to 0 indicate no relationship.
  Results are listed in Table\,\ref{t:lx-fits}. Values obtained for $R$ and $\rho$ suggest that the X-ray
  luminosity of supergiant stars is not correlated with wind parameters. We provide linear fits between
  X-ray luminosity and wind parameters for each luminosity class in Table~\ref{t:lx-fits}.
  For supergiant stars, fitting the X-ray luminosity to a constant gives $\logLx = 32.68 \pm 0.21$. 

This raises the key question on the origin of the $L_{\rm X}$--$L_{\rm 
bol}$ correction for O-stars -- is the X-ray luminosity determined by 
fundamental stellar parameters or by  wind strength? -- a problem 
analogous to the question ``which came first, the chicken or the egg?'.
In the former case, X-rays could be associated with small-scale magnetic fields
on the stellar photosphere \citep[][and references therein]{osk2011}, which in
turn depend on sub-surface structure \citep{cb2011,Urpin2017}.   
The latter case is the well accepted scenario of X-ray generation by strong 
shocks in stellar winds resulting from the line driven instability  
\citep{Feld1997}.   

\input{table_correlations.tex}

As mentioned earlier, these results are based on a limited sample of $35$ stars
  with measured stellar wind velocities. Although these wind velocities are decreasing
  towards later spectral subtypes, there is a significant scatter at each spectral subtype, a scatter also present in the dependence of mass-loss rates on spectral subtype for
  stars of different luminosity classes. Errors on wind velocities and mass-loss rates
  propagate to other wind parameters, such as wind kinetic energy and momentum. A larger sample
  of stars with measured wind velocities and mass-loss rates would help to better constrain
  the correlations between X-ray luminosity and stellar wind parameters.  

For completeness, in Appendix~\ref{sec:appendix3} we show the correlations between
  bolometric luminosity and stellar wind parameters. 

\section{Summary and conclusions}
\label{sec:concl}
The  spectroscopic sample of Galactic O stars, GOSS 
\citep{sotaetal2014, maiz-apellanizetal16-1}, was used to investigate, in a 
homogeneous and consistent way, the X-ray emission from O stars.
We found that 127 stars from GOSS have an unambiguous counterpart in the
  3XMM-DR7 catalog of X-ray sources and are associated to a Gaia-DR2 source.
The stellar parameters of O stars were obtained using spectral 
calibrations presented by \cite{martinsetal05-1}. 
Mass-loss rates of O stars of different spectral types and luminosity classes
were compiled from the literature and used to derive an empirical spectral-type 
mass-loss rate relation. From this empirical relation we derived mass-loss 
rates for all our sample stars. Terminal velocities from \cite{prinjaetal1990} 
  are available for a subsample of 35 of these stars. For this subsample of O stars,
  wind parameters (kinetic luminosity, momentum and density) were
  estimated based on the empirically derived mass-loss rates and the measured terminal velocities.

Interstellar extinction was estimated on the basis of optical photometry
available for all our sample stars.
The stellar distances were estimated on the basis of stellar photometry as 
well using  Gaia DR2. 
The X-ray hardness ratios and fluxes were calculated from count rates 
provided in the 3XMM-DR7 catalog, while X-ray luminosities are based on
dereddened  fluxes and Gaia DR2  parallactic distances.   

We confirm that X-ray luminosities of dwarf and giant O stars correlate with their 
bolometric luminosities. For the subsample of O dwarf and giant stars with
  known terminal velocities we find that their X-ray luminosities are correlated
  with their key wind parameters, such as terminal velocity, 
mass-loss rate, wind kinetic energy, and wind momentum. 
However, these correlations break down for supergiant stars. Furthermore,  O
supergiants have systematically harder X-ray spectra than dwarf and giant stars 
at any given interstellar extinction. We show that the distribution of \LxLbol\
in our sample is non-Gaussian, with the peak of the distribution at 
$\LxLbol\approx -6.6$, minimum value $\LxLbol\approx -7.5$, and a longer tail 
towards higher values up to $\LxLbol\approx -5$. 
We find that the X-ray to bolometric luminosity ratio depends on the spectral 
type: O stars with earlier subtypes are X-ray brighter than with later 
subtypes. We finally find that  \LxLbol\ is independent of whether the stars 
are single or in a binary system, a result in agreement with previous studies.

\begin{acknowledgements}
  We would like to thank the anonymous referee and the editor for their useful comments
  and suggestions that led to the improvement of the manuscript.
The authors are grateful to Dr. W.-R. Hamann for careful reading of the manuscript.
We thank {\em Integrated Activities in 
the High Energy Astrophysics Domain} project for support that enabled this 
work.
This work has been financially supported by the Programme National 
Hautes Energies (PNHE), project 994584 and the Deutsches Zentrum für Luft und 
Raumfahrt (DLR) grant FKZ 50 OR 1508. LMO acknowledges 
partial support by the Russian Government Program of Competitive Growth of 
Kazan Federal University. We thank {\em Integrated Activities in 
the High Energy Astrophysics Domain} project for support that enabled this 
work. We thank {\em Integrated Activities in 
the High Energy Astrophysics Domain} project for support that enabled this 
work. This research has made use of data obtained from the 3XMM XMM-Newton
serendipitous source catalog compiled by the 10 institutes of the XMM-Newton
Survey Science Centre selected by ESA. This research has made use of 
the SIMBAD database, and the VizieR catalog access tool, operated at CDS, 
Strasbourg, France.
\end{acknowledgements}

\bibliographystyle{aa} 
\bibliography{references} %

\begin{appendix}

\section{Soft and hard X-ray to bolometric luminosities}
\label{sec:appendix1}
\begin{figure}[h!]
\includegraphics[width=\linewidth]{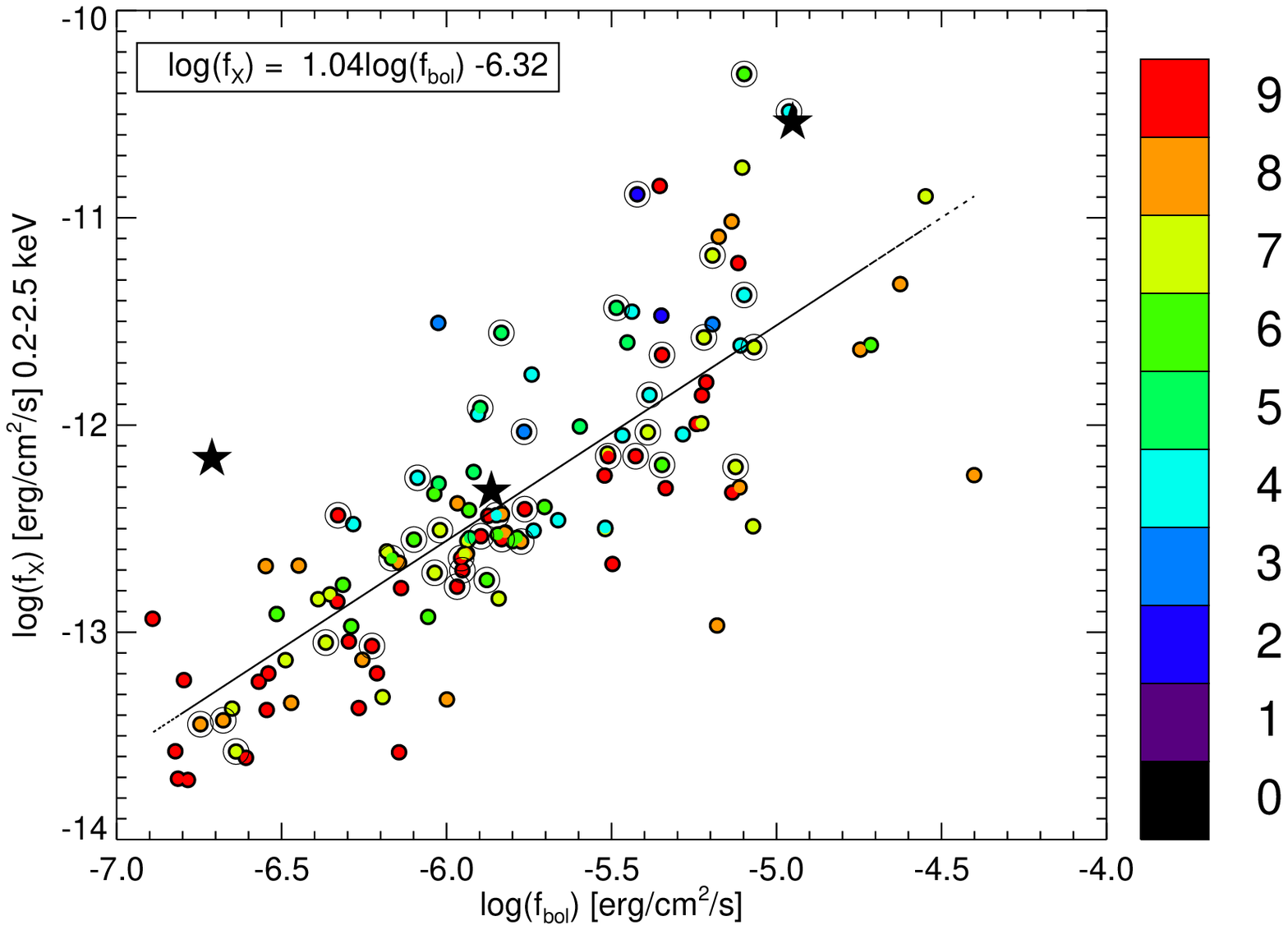}
\includegraphics[width=\linewidth]{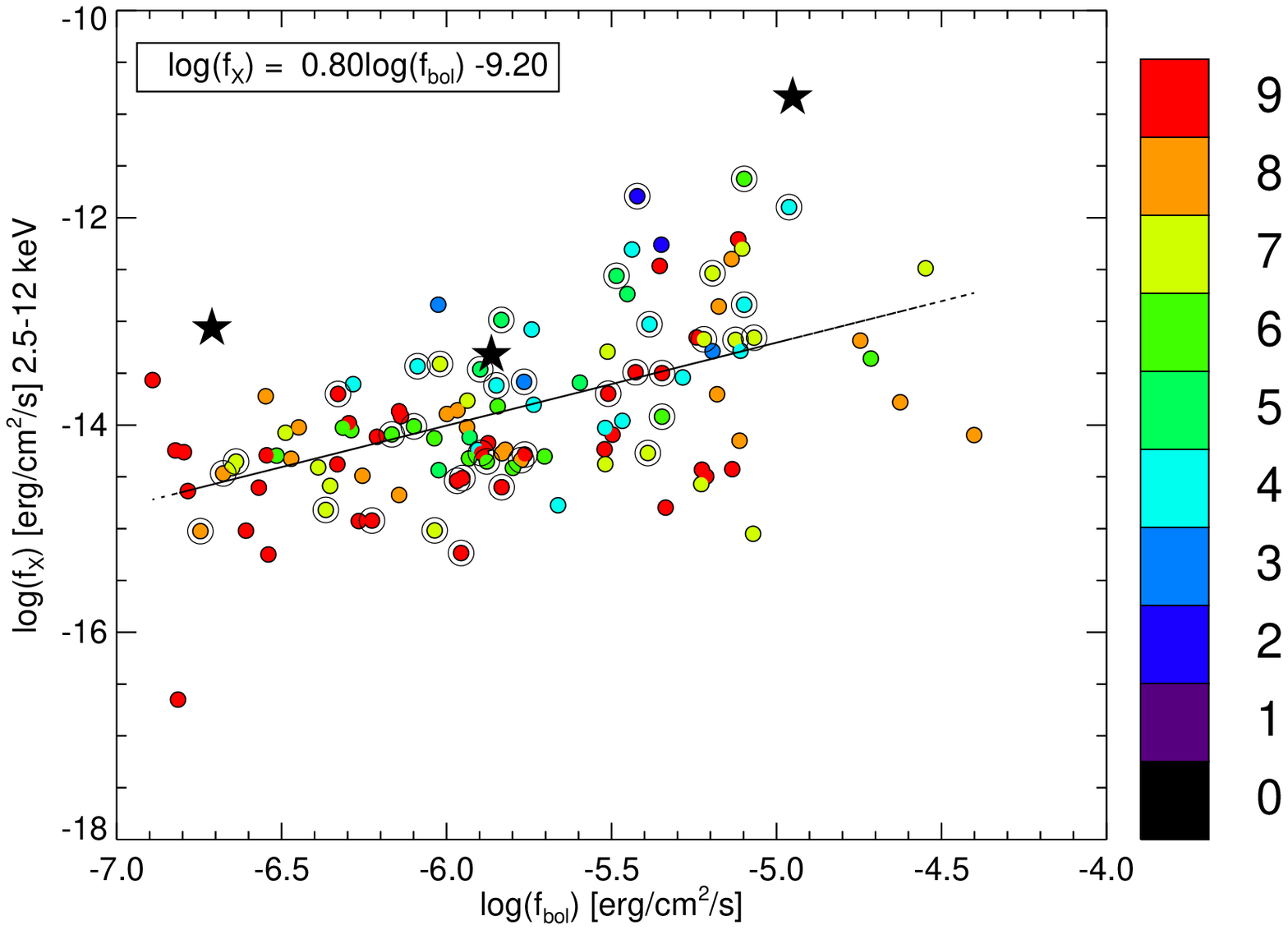}
\caption{\fxfbol\ relation for O stars of all luminosity classes and as a
  function of the spectral subtype (color coded). Known spectroscopic binaries
  are highlighted with a ring around the solid circle and magnetic O stars are
  shown with an asterisk. In the top panel the \fxfbol\ relation in 
  the soft band (0.5\,--\,2.5\,keV) is shown, while in the bottom panel the 
same is shown in  hard X-ray band (2.5\,--\,12\,keV).}
\label{g:fxfbol-SH} 
\end{figure}

For completeness we investigated the \fxfbol\ relation in two different energy 
bands: in soft X-rays (0.5\,--\,2.5\,keV) and in hard X-rays (2.5\,--\,12\,keV).
The correlation between \logfx\ and \logfbol\ is strong in the soft energy band
and moderate in the hard energy band, as shown in Fig.\,\ref{g:fxfbol-SH}
(we note the different scale in the ordinate axis).

\section{Errors in the \LxLbol\ distribution}
\label{sec:appendix2}
\begin{figure}[t!]
\includegraphics[width=\linewidth]{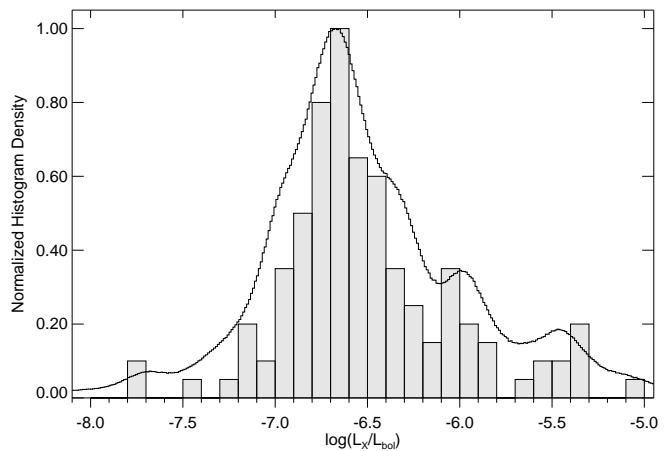}
\caption{\LxLbol\ distribution of our sample stars is shown as a gray histogram and Monte Carlo simulated distribution (see text for details) is shown as a black line.}
\label{fig:hist_LxLbol_errors} 
\end{figure}

To investigate the impact of the estimated errors on the \LxLbol\ distribution we performed a Monte Carlo simulation. For each star we assigned 10\,000 random fluxes normally distributed with mean equal to the measured value and $\sigma$ equal to error. Since for \LxLbol\ we determined upper and lower limits we took as error the maximum difference between these limits and the measured value. We then calculated the normalized histogram of the distribution for these random values. The result is shown in Fig.~\ref{fig:hist_LxLbol_errors} together with the measured distribution. The overall shapes of both distributions are similar, presenting a peak at around the same value and right-skewed. 

\section{Bolometric luminosity as a function of wind parameters}
\label{sec:appendix3}
\begin{figure*}[t!]
\includegraphics[width=0.9\linewidth]{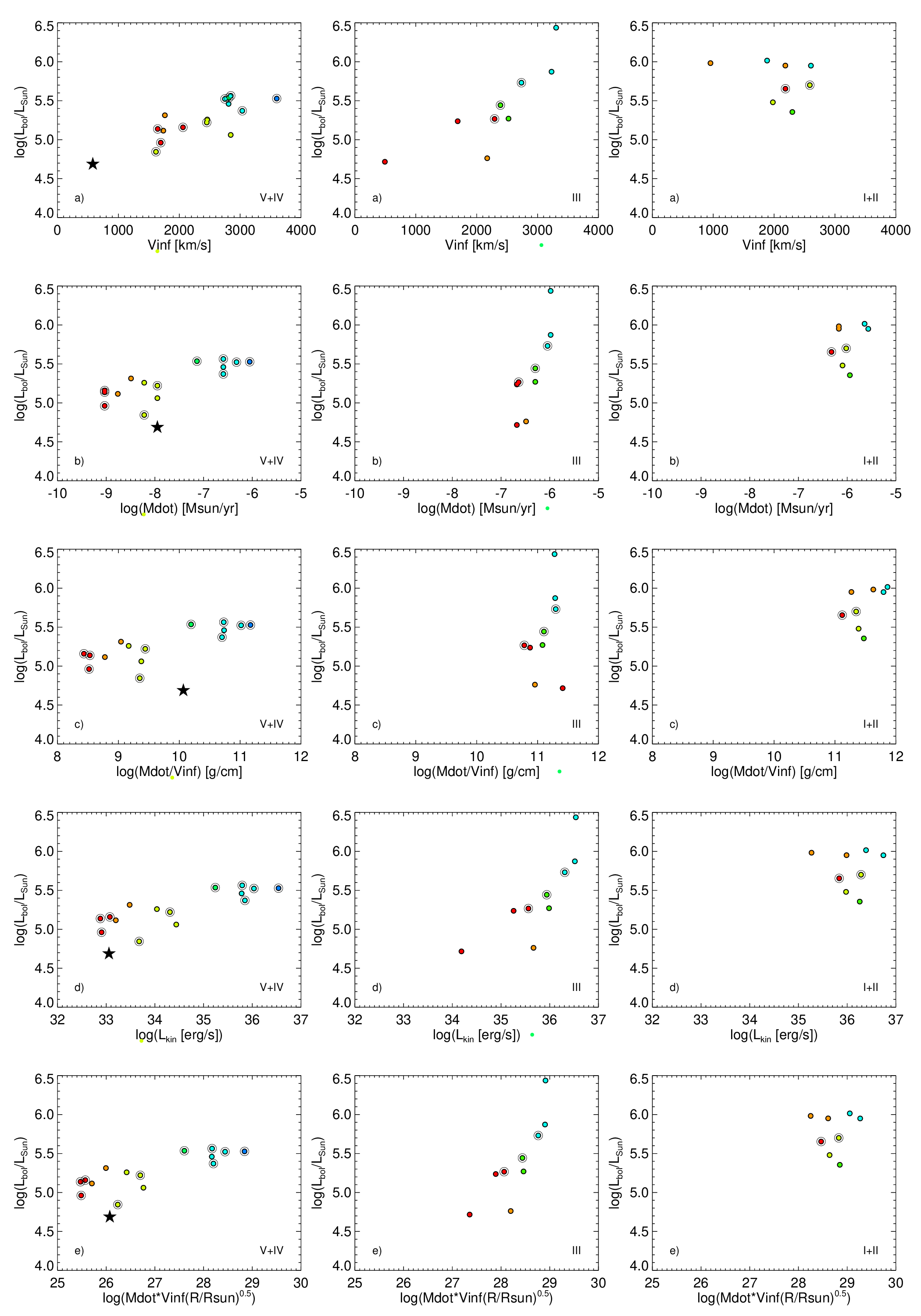}
\caption{Bolometric luminosity against a) terminal wind velocity, b) mass-loss rate
  calculated using the empirical SpT\,--\,\Mdot\ relations derived in
  Sect.~\ref{sec:vinf-massrate}, c) wind density, d) kinetic wind luminosity and
  e) modified stellar wind momentum for different luminosity
  classes in each panel. Symbols have been color coded according to spectral type.}\label{fig:Ekin_Lbol_Gaia} 
\end{figure*}

\begin{figure*}[t!]
\includegraphics[width=0.9\linewidth]{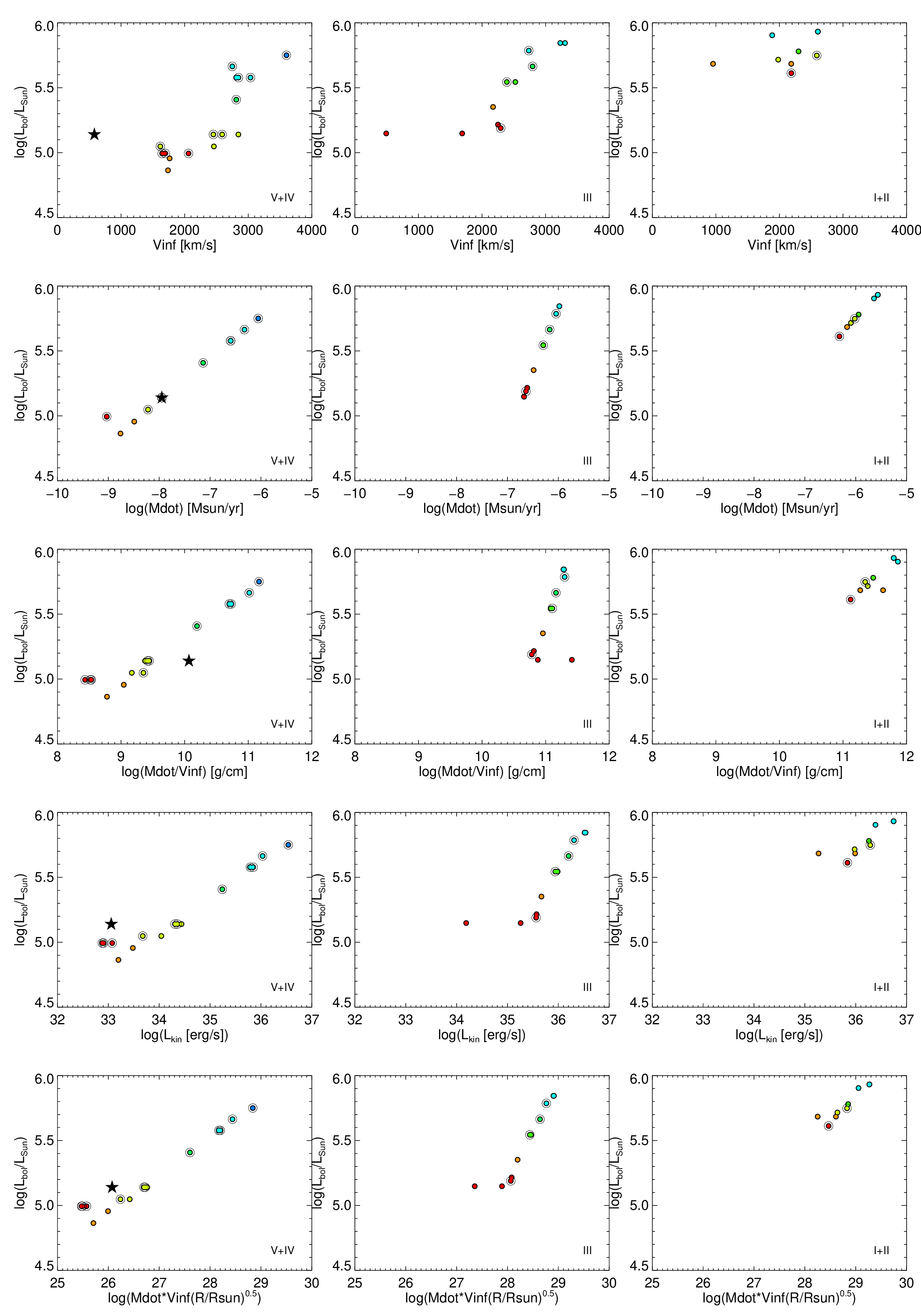}
\caption{Bolometric luminosity against a) terminal wind velocity, b) mass-loss rate
  calculated using the empirical SpT\,--\,\Mdot\ relations derived in
  Sect.~\ref{sec:vinf-massrate}, c) wind density, d) kinetic wind luminosity and
  e) modified stellar wind momentum for different luminosity
  classes in each panel. Symbols have been color coded according to spectral type.}\label{fig:Ekin_Lbol} 
\end{figure*}

The bolometric luminosity is shown as a function of wind parameters in 
Figs.\,\ref{fig:Ekin_Lbol_Gaia} and \ref{fig:Ekin_Lbol}, as a function of luminosity class (dwarfs and 
subdwarfs in the left panels, giants in the middle panels, and supergiants in the 
right panels). While dwarf and giant stars stars show a moderate correlation 
with the terminal velocity, supergiant stars do not. 
Wind kinetic luminosity, 
momentum, and density are dependent on \Vinf\ and \Mdot. Therefore, the 
correlations with \Lbol\ are reflecting a correlation with these two 
fundamental 
parameters.  
While in Fig.\,\ref{fig:Ekin_Lbol_Gaia} luminosities have been calculated 
based on the
parallactic distances \citep{BJ2018}, in 
Fig.\,\ref{fig:Ekin_Lbol} luminosities are determined using the tabulated values 
by \cite{martinsetal05-1} for different spectral types and luminosity classes.

\end{appendix}

\end{document}

%% file: table_catalog.tex
\begin{table*}
  \caption{An example of  entry to the \xo\ catalog and column descriptions.}
\label{t:cat}
\begin{center}
\small
\begin{tabular}{lc|r|l}
\hline
\hline
Column Name             &            &                         & Explanation               \\
\hline
GOSS                    &            &  006.01-01.20\_01       & Name in the GOSS catalogue \\
NAME                    &            &  9 Sgr                  & Alternative name \\
XMM-NAME                &            &  3XMM J180352.4-242138  & 3XMM Name as in DR7 \\
GaiaDR2\_source\_id &           &  4066022591147527552    & Gaia source identifier, identical to the Gaia-DR2 source\_id.\\
RA                      &            &  18 03 52.446           & RAJ2000 \\
DEC                     &            &  -24 21 38.64           & DecJ2000 \\
\Mv\                    & [mag]      & -5.5                    & Absolute magnitude from \cite{martinsetal05-1}     \\ 
error\_\Mv        & [mag]       & 0.29                    & Absolute magnitude error \\ 
\Mbol                   & [mag]       & -9.41                   & Bolometric magnitude from \cite{martinsetal05-1}    \\
e\_\Mbol          & [mag]       &  0.43                  & Bolometric magnitude error    \\
$ T_{\rm eff}$           & [K]        &  43000                  & Effective temperature from \cite{martinsetal05-1} \\
e\_T$_{\rm eff}$   & [K]        &  2000                  & Effective temperature error\\
$\log{g}$               &            &  3.92                   & Effective gravity from \cite{martinsetal05-1} \\
$e\_\log{g}$            &            &  0.1                   & Effective gravity error\\
$\log{L_{\rm bol,S}}$     & [erg\,s$^{-1}$]      & 5.66  & Bolometric luminosity derived from spectrophotometric distance\\
\Vinf\                  & [km\,s$^{-1}$]         &  2750         & Terminal wind velocity from \cite{prinjaetal1990}\\
$\log\Mdot_{\rm Vink}$  & [$M_\odot$\,yr$^{-1}$]  & -5.62         & Mass-loss rate from the \cite{vinketal2000} recipe\\
$\log\Mdot$             & [$M_\odot$\,yr$^{-1}$]  & -6.32        & Mass-loss rate from this work\\
distance\_Gaia         & [kpc]                       & 1.2      & Parallactic distance from \cite{BJ2018} \\
min\_distance\_Gaia    & [kpc]                       & 1.1      & Minimum parallactic distance from \cite{BJ2018} \\
max\_distance\_Gaia    & [kpc]                       & 1.3      & Maximum parallactic distance from \cite{BJ2018} \\
distance\_\Mv          & [kpc]                       & 1.4      & Spectrophotometric distance \\
error\_distance\_\Mv   & [kpc]                       & 0.1      & Error on the spectrophotometric distance \\
$B$              & [mag]                       & 5.94     & B magnitude \\
e\_$B$             & [mag]                       & 0.01     & B magnitude error\\
$V$                     & [mag]                      & 5.96     & V magnitude          \\
e\_$V$              & [mag]                      & 0.01     & V magnitude error      \\
\EBV                   & [mag]                      & 0.26     & Color excess            \\
e\_\EBV         & [mag]                      & 0.02     & Color excess error           \\
$\Av$             & [mag]                      & 0.81     & Extinction in the V band\\
e\_$\Av$          & [mag]                      & 0.08     & Error in the extinction in the V band\\
$N_{\rm H}$              & [cm$^{-2}$]                 & 1.5E21   & Hydrogen column density \\
PN count\_rate          & [s$^{-1}$]                  & 1.284    & PN Count rate [0.2-12 keV] \\
error\_count\_rate      &  [s$^{-1}$]                 & 0.017    & PN Count rate error [0.2-12 keV]\\
$f_{\rm X}$              &[erg\,s$^{-1}$\,cm$^{-2}$]   & 4.4E-12  & Unabsorbed X-ray flux [0.2-12 keV]\\
$f_{\rm X,min}$          &  [erg\,s$^{-1}$\,cm$^{-2}$]  & 4.3E-12  & Minimum X-ray flux [0.2-12 keV] \\
$f_{\rm X,max}$          & [erg\,s$^{-1}$\,cm$^{-2}$]   & 4.4E-12  & Maximum X-ray flux [0.2-12 keV] \\
$\log(\fbol)$    & [erg\,s$^{-1}$\,cm$^{-2}$]   & -5.098    & Bolometric flux \\
e$\_\log(\fbol)$ & [erg\,s$^{-1}$\,cm$^{-2}$]   & 0.032    & Bolometric flux error\\
\fxfbol                &                             & -6.26    & X-ray to bolometric flux ratio \\
min\_\fxfbol     &                             & -6.30    & Minimum X-ray to bolometric flux ratio \\
max\_\fxfbol     &                             & -6.22    & Maximum X-ray to bolometric flux ratio \\
$\log{L_{\rm X}}$       & [erg\,s$^{-1}$]              & 32.84   & X-ray luminosity [0.2-12 keV]\\
$\log{L_{\rm X,min}}$    & [erg\,s$^{-1}$]             & 32.74    & minimum X-ray luminosity [0.2-12 keV]\\
$\log{L_{\rm X,max}}$    & [erg\,s$^{-1}$]             & 32.95    & maximum X-ray luminosity [0.2-12 keV]\\
$\log{L_{\rm bol,P}}$    & [erg\,s$^{-1}$]             & 39.10    & Bolometric luminosity derived from parallactic distance\\
min\_$\log{L_{\rm bol,P}}$ & [erg\,s$^{-1}$]           & 38.98    & Minimum bolometric luminosity \\
max\_$\log{L_{\rm bol,P}}$ & [erg\,s$^{-1}$]           & 39.24    & Maximum bolometric luminosity \\
HR2                    &                             & -0.493   & Hardness ratio HR2           \\
error\_HR2             &                             & 0.013    & Error on the hardness ratio HR2 \\
HR3                    &                             & -0.814   & Hardness ratio HR3        \\
error\_HR3             &                             & 0.017    & Error on the hardness ratio HR3 \\
\hline
\end{tabular}
\end{center}
\end{table*}

%% file: table_mass-loss-rates.tex
\begin{table}
\begin{center}
\caption{Stellar mass-loss rates of O type stars collected 
from the literature.  An interested reader is strongly encouraged to consult 
original papers for details of the spectroscopic analyses.  }
\label{t:mass-rates}
\begin{tabular}{lccc}
\hline
\hline
HD Name & SpT & \Mdot\ [$M_\odot$\,yr$^{-1}$] & Ref. \\
\hline
\multicolumn{4}{c}{Luminosity class I}  \\
\hline
HD 66811   & O4I    & 2.5e-6   & 1 \\
HD 16691   & O4I    & 3.0e-6   & 2 \\
HD 190429A & O4If   & 2.1e-6   & 2 \\
HD 15570   & O4I    & 2.8e-6   & 1 \\
HD 14947   & O4.5If & 2.8e-6   & 1 \\
HD 210839  & O6.5I  & 1.6e-6   & 1 \\
HD 163758  & O6.5I  & 1.6e-6   & 2 \\
HD 192639  & O7.5I  & 1.3e-6   & 1 \\
\hline
\multicolumn{4}{c}{Luminosity class II--III}  \\
\hline
HD 93250    & O4III     & 5.6e-7   & 3 \\
HD 15558    & O4.5III   & 1.9e-6   & 4 \\
HD 190864   & O6.5III   & 4.6e-7   & 4 \\
HD 34656    & O7II      & 3.0e-7     & 4 \\
HD 24912    & O7.5III   & 4.0e-7   & 4 \\
HD 36861    & O8III     & 3.0e-7   & 4 \\
$\delta$\,Ori\,Aa1 & O9.5II    & 4.0e-7     & 5 \\
\hline
\multicolumn{4}{c}{Luminosity class V}  \\
\hline
HD 46223   & O4V       & 3.2e-7   & 3  \\
HD 15629   & O4.5V     & 3.2e-7   & 3  \\
HD 93204   & O5.5V     & 1.8e-7   & 3  \\
HD 42088   & O6V((f))z & 1.0e-8   & 3  \\
HD 152590  & O7.5Vz    & 1.6e-8   & 3  \\
HD 93028   & O9IV      & 1.0e-9   & 3  \\
HD 46202   & O9.2V     & 1.3e-9   & 3  \\
HD 38666   & O9.5V     & 3.0e-10  & 3  \\
HD 34078   & O9.5V     & 3.0e-10  & 3  \\
HD 36512  & O9.7V     & 5.0e-10  & 6 \\

\hline
\end{tabular}
\tablebib{
(1) \citet{Surlan2013};
(2) \cite{bouretetal12-1}; 
(3) \citet{martinsetal05-1}; 
(4) \citet{repolustetal04-1}, note that the mass-loss rates in the table  are 
reduced by a facor of 3; 
(5) \citet{shenaretal15-1}. 
(6) \citet{shenar2017}}
\end{center}
\end{table}

%% file: table_fits.tex
\begin{table}
\caption{Number of O stars, correlation coefficient $R$ (see text for its 
definition), and fit coefficients for 
\mbox{$\logfx\,=\,a\,+\,b\times\logfbol$}.}\label{t:fitparams}
\begin{center}
\begin{tabular}{c|c|c|c|c}
\hline
\hline
Lum. Class & \# & $R$ & $a$ & $b$  \\
\hline
All  &  127 & 0.78 &  $-6.21  \pm  0.44$ & $1.05  \pm  0.08$\\
V    &   60 & 0.80 &  $-7.05  \pm  0.54$ & $0.92  \pm  0.09$\\
IV   &   14 & 0.83 &  $-7.92  \pm  0.94$ & $0.78  \pm  0.16$\\
V+IV &   74 & 0.80 &  $-7.15  \pm  0.47$ & $0.90  \pm  0.08$\\
III  &   21 & 0.89 &  $-3.91  \pm  1.03$ & $1.45  \pm  0.18$\\
I+II &   30 & 0.55 &  $-6.19  \pm  1.63$ & $1.04  \pm  0.30$\\

\hline
\end{tabular}
\end{center}
\end{table}

%% file: table_stats.tex
\begin{table}
\caption{Mean \fxfbol.}\label{t:meanvalues}
\begin{center}
\begin{tabular}{c|c}
\hline
\hline
Lum. Class & \fxfbol  \\
\hline
All         & $-6.50  \pm  0.47$ \\
V           & $-6.56  \pm  0.39$ \\
IV          & $-6.61  \pm  0.25$ \\
V+IV        & $-6.57  \pm  0.37$ \\
III         & $-6.51  \pm  0.42$ \\
I+II        & $-6.40  \pm  0.62$ \\
SpT early   & $-6.23  \pm  0.44$ \\
SpT late    & $-6.62  \pm  0.43$ \\
Single      & $-6.54  \pm  0.47$ \\
Binaries    & $-6.41  \pm  0.46$ \\
\hline
\end{tabular}
\end{center}
\end{table}

%% file: table_correlations.tex
\begin{table}
  \caption{Pearson (R) and Spearman's Rank ($\rho$) correlation coefficients (see text for descriptions) and fit coefficients for each luminosity class (LC). }\label{t:lx-fits}
\begin{center}
\begin{tabular}{c|r|r|r|r}
\hline\hline
LC & $R$ & $\rho$ & $a$ & $b$ \\
\hline
\multicolumn{5}{c}{Terminal velocity} \\
\multicolumn{5}{c}{$\log\Lx = a+b\times\Vinf$} \\
\hline
All   & $0.5$  & 0.6 & $  30.7  \pm  0.6$  & $  (8 \pm 2)\times10^{-4}$ \\
V     & $0.5$  & 0.6 & $  31.1  \pm  0.7$  & $  (6 \pm 3)\times10^{-4}$ \\
III   & $0.8$  & 0.8 & $  28.9  \pm  1.4$  & $  (10\pm 5)\times10^{-4}$ \\
I     & $0.2$  & 0.2 & $  31.6  \pm  2.2$  & $  (5 \pm10)\times10^{-4}$ \\
\hline
\multicolumn{5}{c}{Mass-loss rate} \\
\multicolumn{5}{c}{$\log\Lx = a+b\times\log\Mdot$} \\
\hline
All & $  0.4$  & 0.5 & $  34.3  \pm  0.8$ & $   0.3  \pm  0.1$ \\
V   & $  0.4$  & 0.6 & $  34.7  \pm  1.4$ & $   0.3  \pm  0.2$ \\
III & $  0.7$  & 0.8 & $  49.5  \pm  6.8$ & $   2.7  \pm  1.1$ \\
I   & $ -0.1$  & -0.1& $  31.8  \pm  6.0$ & $  -0.2  \pm  1.0$ \\
\hline
\multicolumn{5}{c}{Kinetic luminosity} \\
\multicolumn{5}{c}{$\log\Lx = a+b\times\log\Lkin$} \\
\hline
All  & $ 0.4$ & 0.6 & $   23.6  \pm  3.8$ & $  0.3  \pm  0.1$ \\
V    & $ 0.4$ & 0.6 & $   23.7  \pm  5.0$ & $  0.3  \pm  0.1$ \\
III  & $ 0.7$ & 0.7 & $  -28.0  \pm 22.8$ & $  1.7  \pm  0.6$ \\
I    & $ 0.0$ & 0.3 & $   31.8  \pm 32.7$ & $  0.0  \pm  0.9$ \\
\hline
\multicolumn{5}{c}{Wind density} \\
\multicolumn{5}{c}{$\log\Lx = a+b\times\log N_1$} \\
\hline
All  & $ 0.3$ & 0.4 & $   30.1  \pm  1.3$ & $  0.2  \pm  0.1$ \\
V    & $ 0.4$ & 0.6 & $   29.5  \pm  1.9$ & $  0.3  \pm  0.2$ \\
III  & $ 0.7$ & 0.7 & $   -5.9  \pm 18.4$ & $  3.5  \pm  1.7$ \\
I    & $-0.1$ & -0.2& $   35.6  \pm 11.6$ & $ -0.3  \pm  1.0$ \\
\hline
\multicolumn{5}{c}{Wind momentum}  \\
\multicolumn{5}{c}{$\log\Lx = a+b\times\log D$}  \\
\hline
All  & $ 0.4$ & 0.5  & $   25.9  \pm  3.0$ & $  0.2  \pm  0.1$ \\
V    & $ 0.4$ & 0.6  & $   25.3  \pm  4.2$ & $  0.3  \pm  0.2$ \\
III  & $ 0.7$ & 0.7  & $  -24.2  \pm 21.9$ & $  2.0  \pm  0.8$ \\
I    & $-0.0$ & -0.1 & $   34.3  \pm 29.0$ & $ -0.1  \pm  1.0$ \\
\hline\hline
\end{tabular}
\end{center}
\end{table}